\documentclass[aip,jcp,reprint,amsmath,amssymb,nobalancelastpage]{revtex4-2}
\usepackage{graphicx}
\usepackage[suffix=]{epstopdf}
\usepackage{natmove}
\newcommand{\vv}[1]{\mathbf{#1}}
\renewcommand{\d}[1]{\ensuremath{\operatorname{d}\!{#1}}}

\begin{document}
\title{Exploring the role of hydrodynamic interactions in spherically-confined drying colloidal suspensions}

\author{Mayukh Kundu}
\thanks{These authors contributed equally.}
\affiliation{Department of Chemical Engineering, Auburn University, Auburn, AL 36849, USA}

\author{Kritika Kritika}
\thanks{These authors contributed equally.}
\affiliation{Leibniz-Institut f{\"u}r Polymerforschung Dresden e.V., Hohe Stra{\ss}e 6, 01069 Dresden, Germany}
\affiliation{Institut f{\"u}r Theoretische Physik, Technische Universit{\"a}t Dresden, 01069 Dresden, Germany}

\author{Yashraj M. Wani}
\affiliation{Institute of Physics, Johannes Gutenberg University Mainz, Staudingerweg 7, 55128 Mainz, Germany}

\author{Arash Nikoubashman}
\email{anikouba@ipfdd.de}
\affiliation{Leibniz-Institut f{\"u}r Polymerforschung Dresden e.V., Hohe Stra{\ss}e 6, 01069 Dresden, Germany}
\affiliation{Institut f{\"u}r Theoretische Physik, Technische Universit{\"a}t Dresden, 01069 Dresden, Germany}

\author{Michael P. Howard}
\email{mphoward@auburn.edu}
\affiliation{Department of Chemical Engineering, Auburn University, Auburn, AL 36849, USA}

\begin{abstract}
We study the distribution of colloidal particles confined in drying spherical droplets using both dynamic density functional theory (DDFT) and particle-based simulations. In particular, we focus on the advection-dominated regime typical of aqueous droplets drying at room temperature and systematically investigate the role of hydrodynamic interactions during this nonequilibrium process. In general, drying produces transient particle concentration gradients within the droplet in this regime, with a considerable accumulation of particles at the droplet's liquid--vapor interface. We find that these gradients become significantly larger with pairwise hydrodynamic interactions between colloidal particles instead of a free-draining hydrodynamic approximation; however, the solvent's boundary conditions at the droplet's interface (unbounded, slip, or no-slip) do not have a significant effect on the particle distribution. DDFT calculations leveraging radial symmetry of the drying droplet are in excellent agreement with particle-based simulations for free-draining hydrodynamics, but DDFT unexpectedly fails for pairwise hydrodynamic interactions after the particle concentration increases during drying, manifesting as an ejection of particles from the droplet. We hypothesize that this unphysical behavior originates from an inaccurate approximation of the two-body density correlations based on the bulk pair correlation function, which we support by measuring the confined equilibrium two-body density correlations using particle-based simulations. We identify some potential strategies for addressing this issue in DDFT.
\end{abstract}

\maketitle

\section{Introduction}
Spray \cite{walton:1999} and emulsion \cite{velev:1996, rosca:2004} drying can produce nearly spherical droplets comprised of a suspension of colloidal particles. These particles, confined inside the droplet, can subsequently assemble into larger supraparticles as the droplets dry \cite{wintzheimer:2018, bassani:acsnano:2024}. The typical diameter of a supraparticle is on the order of 10 $\mu$m to 100 $\mu$m, depending on the initial droplet volume, the initial particle concentration, and the particle size, among other process variables. Supraparticles can exhibit emergent properties beyond those of their constituent particles, making them a versatile materials platform with applications in photonics \cite{iskandar:2003, wang:2020, luo:2014, patil:aom:2022, heil:scadv:2023}, catalysis\cite{leon:2020, hou:2020}, drug delivery \cite{bodmeier:1987}, and fertilizers \cite{de_lima:2023}. Many of these properties are dependent on the final, potentially inhomogeneous arrangement of one or more types of particles inside the supraparticle\cite{wintzheimer:2018, liu:march:2019, liu:december:2019, liu:2022, heil:scadv:2023}, so it is important to be able to predict the particle distribution in a supraparticle from process variables that can be engineered.

Computer simulations can be powerful tools for understanding how particles assemble during drying. However, drying is a complicated nonequilibrium process, involving thermodynamic and transport phenomena across multiple length and time scales, and there are accordingly various methodologies that can be used to model it\cite{routh:2004, howard:lng:2017, tang:jcp:2019}. These methodologies can be broadly categorized as either particle-based or continuum. In particle-based methods, the colloidal particles are represented explicitly and typically interact with each other through (pairwise) potential-energy functions representing their colloidal forces, while the solvent is included explicitly as additional smaller particles\cite{yamaguchi:1998, chen:2013}, in a simplified fashion by coarse-graining \cite{chen:1998, malevanets:1999}, or implicitly through the particles' equations of motion\cite{brady:jfm:1988}. In practice, particle-based methods are computationally limited in the length and time scales that they can describe by the number of particles and the number of numerical time-integration steps required. Continuum models overcome these limitations by representing the average local density of colloidal particles rather than the particles themselves \cite{routh:2004, routh:2013, sear:pre:2017, howard:lng:2017, zhou:prl:2017, chun:2020, he:lng:2021, rees-zimmerman:2021, yoo:2022, kundu:2022}. The particle densities evolve according to a conservation equation with flux models constructed to reasonably capture smaller-scale effects, including those from the particle interactions and solvent, that have been averaged over \cite{brady:2011bh}.

Both particle-based and continuum models hence require making a choice about how to model the solvent and its effects on the suspended particles, but this choice can present unexpected challenges. Previous studies have shown that solvent-mediated hydrodynamic interactions (HIs) between colloidal particles can play a major role in drying-induced structure formation \cite{antonia:2018, howard:jcp:2018, howard:jcp:2020, liu:march:2019, chun:2020}. For example, models with free-draining HIs, which neglect the hydrodynamic couplings between particles, are theoretically expected to incorrectly predict the extent of drying-induced stratification in multicomponent suspensions \cite{sear:pre:2017}. Liu et al.~found evidence of this behavior in drying droplets containing a bidisperse colloidal suspension, where simulations without HIs produced much more pronounced core--shell supraparticles than observed in experiments\cite{liu:march:2019}.  A similar effect was noted in simulations of drying mixtures of long and short polymers, where neglecting HIs between polymers led to stratification that did not occur when such HIs were included \cite{antonia:2018, howard:jcp:2020}. Howard et al.~also showed that HIs influenced the kinetics of evaporation-induced colloidal crystallization in thin films \cite{howard:jcp:2018}. Conversely, though, some studies have indicated that HIs may not always play an important role for structure formation during drying. Tang et al.~found no significant differences in the stratification of a drying coating containing a bidisperse colloidal suspension using explicit-solvent simulations (with HIs between particles) or implicit-solvent simulations with only free-draining HIs\cite{tang:jcp:2019}. Similarly, Yetkin et al.~found that HIs between particles had only minimal impact on the overall structure formation in drying droplets containing ellipsoidal nanoparticles\cite{yetkin:2024}. These conflicting observations highlight the importance of further investigating the role of HIs in drying colloidal suspensions so that appropriate modeling choices can be made.

\begin{figure}
    \includegraphics{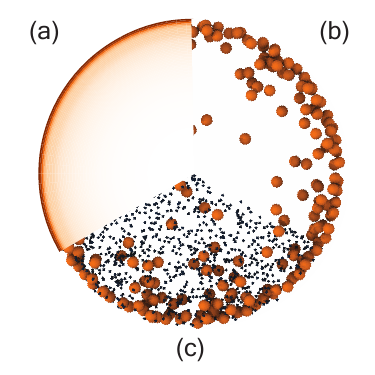}
    \caption{Schematic of the different simulation models employed: (a) dynamic density functional theory (DDFT), (b) Brownian dynamics (BD), and (c) multiparticle collision dynamics (MPCD).}
    \label{fig:schematic}
\end{figure}

In this work, we use both continuum and particle-based models to simulate suspensions of hard-sphere colloidal particles confined within a drying droplet, both without and with the presence of pairwise HIs between particles (Fig.~\ref{fig:schematic}). The continuum models [Fig.~\ref{fig:schematic}(a)] are based on dynamic density functional theory (DDFT), which provides a structured framework for formulating approximate continuum-scale flux models from particle-level interactions and dynamics \cite{marconi:jcp:1999, rex:prl:2008, rex:epje:2009}. DDFT has previously been used to model drying-induced stratification in bidisperse suspensions, but these models did not consider HIs between particles\cite{howard:lng:2017, howard:lng:2017B, zhou:prl:2017, he:lng:2021, kundu:2022}. Here, we derive an expression for the particle flux due to pairwise, unconfined HIs in a dilute suspension assuming radial symmetry in the droplet. We then analyze and discuss the challenges associated with extending this approach to nondilute suspensions. To test our DDFT method and to further probe the importance of HIs, we perform complementary particle-based simulations. The first particle-based model, Brownian dynamics (BD) [Fig.~\ref{fig:schematic}(b)], initially uses a free-draining hydrodynamic approximation that we subsequently extend to incorporate pairwise HIs, assuming unbounded flow across the droplet's liquid--vapor interface. To assess the impact of the solvent boundary conditions at the droplet interface, we also develop a mesoscale model based on the multiparticle collision dynamics (MPCD) method that incorporates (no-)slip boundary conditions on the solvent at the droplet's interface [Fig.~\ref{fig:schematic}(c)]. Overall, this work provides insights into the role of HIs in nonequilibrium, inhomogeneous, and confined colloidal suspensions, highlighting key factors that must be carefully considered when simulating these systems at both particle and continuum scales.

The rest of this article is organized as follows. In Sec.~\ref{sec:methods}, we describe the model considered for the spherically-confined drying colloidal suspension and the three different simulation methods (BD, MPCD, and DDFT) employed. In Sec.~\ref{sec:results}, we compare the particle distribution in the drying droplet that is generated by the different simulation methods assuming either free-draining or pairwise HIs between colloidal particles. We conclude in Sec.~\ref{sec:conclusions} with a summary of our main findings and forward-looking comments on paths to further improve the accuracy of models for drying colloidal suspensions.

\section{Model and Methods}
\label{sec:methods}
We studied a colloidal suspension of hard-sphere particles with diameter $d$ (radius $a = d/2$) confined in a spherical droplet with radius $R$. To model drying, the droplet radius decreased as a function of time $t$ according to
\begin{equation}
    R(t) = \sqrt{R_0^2-\frac{\alpha}{4\pi}t},
    \label{eq:interface}
\end{equation}
where $\alpha$ is the constant rate of change of the droplet's surface area. This form of $R(t)$ was derived by Langmuir based on mass-transfer considerations for a spherical liquid droplet surrounded by its own vapor \cite{langmuir:1918}. The parameter $\alpha$ depends on physical properties such as the solvent's vapor pressure and diffusivity, but here we treat it as an adjustable model parameter that controls the rate and qualitative features of drying. We characterize the drying conditions using the dimensionless P\'{e}clet number ${\rm Pe} = R_0 |V_0| / D_0$, which is the ratio of the typical diffusion and drying times. In this definition, $V_0 = -\alpha/(8\pi R_0)$ is the initial velocity of the receding droplet interface, and $D_0 = k_{\rm B} T/(6 \pi \mu a)$ is the translational self-diffusion coefficient of a spherical particle with no-slip boundary conditions in a solvent with viscosity $\mu$ at temperature $T$, where $k_{\rm B}$ is the Boltzmann constant. When ${\rm Pe} \ll 1$, the drying process is diffusion-dominated and particles are expected to adopt their equilibrium distribution in the droplet. However, when ${\rm Pe} \gg 1$, the drying process is advection-dominated, and nonequilibrium concentration gradients are expected to develop. Given that equilibrium models for confined colloidal suspensions are well studied\cite{Roth:2010ei, reinhardt:2012}, we chose to focus this work on modeling the less-explored advection-dominated drying regime\cite{routh:2004, routh:2013, schulz:sm:2018} and selected $\alpha$ such that ${\rm Pe} \approx 10$, which corresponds to the conditions typically encountered for aqueous droplets drying at room temperature\cite{liu:march:2019}.

Particle interactions with the droplet's liquid--vapor interface were modeled by a repulsive harmonic potential\cite{howard:lng:2017, liu:december:2019}
\begin{equation}
    \beta \psi(\vv{x},t) = \begin{cases}
\displaystyle \frac{\kappa}{2}\left[|\vv{x}|-R(t)+a\right]^2,& |\vv{x}| > R(t)-a \\
0,& {\rm otherwise}
\end{cases},
\label{eq:slv}
\end{equation}
where $\beta = 1/(k_{\rm B}T)$, $\vv{x}$ is the position of the particle, and $\kappa$ is a spring constant. We used $\kappa = 100\,d^{-2}$ to ensure the particles remained essentially fully immersed in the droplet. This model does not take into consideration capillary forces between particles at the interface and assumes the droplet retains a spherical shape (i.e., our model does not capture buckling)\cite{bahadur:2011}. It also neglects effects of skin formation on the rate of drying \cite{wintzheimer:2018}.

The initial average volume fraction of particles in the droplet was $\eta_0 = N (a/R_0)^3$, where $N$ is the constant number of particles. The average volume fraction of particles in the droplet $\eta$ increases according to $\eta = \eta_0 (R_0/R)^3$ as $R$ decreases during drying. We considered initial volume fractions of $\eta_0 = 0.05$, 0.10, and 0.20 in a droplet with initial radius $R_0 = 25\,d$, giving between 6250 and 25000 particles in the droplet. We dried the droplet until $\eta = 0.5$ regardless of $\eta_0$. At this volume fraction, we expect the particle motion to slow considerably, such that significant structural changes are no longer occurring due to diffusion.

\subsection{Particle-based simulations}
\subsubsection{Brownian dynamics}
\label{sec:methods:bd}
In our BD simulations, we approximated hard-sphere interactions between particles using the purely repulsive Weeks--Chandler--Andersen (WCA) potential\cite{weeks:1971},
\begin{equation}
\beta u(r) = \begin{cases}
\displaystyle 4 \left[\left(\frac{d}{r}\right)^{12}-\left(\frac{d}{r}\right)^6\right] + 1,& r \le 2^{1/6} d \\
0,& {\rm otherwise}
\end{cases},
\label{eq:wca}
\end{equation}
where $r$ is the distance between the centers of two particles. The displacements $\Delta X$ of the $N$ particle positions $X = ({\vv x}_1, ..., {\vv x}_N)$, where $\vv{x}_i$ is the position of particle $i$, during a time step $\Delta t$ were \cite{batchelor_green:jfm:1972, batchelor:jfm:1983, brady:arfm:1988,russel:1989, happel:2012}
\begin{equation}
\Delta X = [\mathcal{M} \cdot \mathcal{F} + k_{\rm B} T(\nabla_X\cdot\mathcal{M})]\Delta t + \Delta \mathcal{W},
\label{eq:bd}
\end{equation}
where $\mathcal{M}$ is the grand mobility tensor coupling particle forces to translational velocities, $\mathcal{F}$ is the vector of forces acting on the particles [due to Eqs.~\eqref{eq:slv} and \eqref{eq:wca}], and $\Delta \mathcal{W}$ is the vector of random particle displacements that is Gaussian-distributed with zero mean $\langle \Delta \mathcal{W} \rangle = \vv{0}$ and covariance $\langle \Delta\mathcal{W} \Delta \mathcal{W} \rangle = 2 k_{\rm B} T \mathcal{M} \Delta t$. The divergence of $\mathcal{M}$ is taken with respect to $X$.

We employed two approximations of $\mathcal{M}$: a free-draining approximation, in which particles individually experienced drag from the solvent, and a pairwise far-field approximation, in which particles were also hydrodynamically coupled to each other through the Rotne--Prager--Yamakawa (RPY) tensor\cite{rotne:1969, yamakawa:1970}. Specifically, the elements of $\mathcal{M}$ were expressed as pairwise tensors $\vv{M}^{(ij)}$ relating the velocity of particle $i$ to the force on particle $j$. In the free-draining approximation, $\vv{M}^{(ii)} = \vv{M}^{(1)}$, where
\begin{equation}
    \vv{M}^{(1)} = \frac{1}{6\pi \mu a}\vv{I}
\end{equation}
and $\vv{I}$ is the identity tensor, and $\vv{M}^{(ij)} = \vv{0}$ for $i \ne j$. In the RPY approximation, $\vv{M}^{(ii)} = \vv{M}^{(1)}$ as well but $\vv{M}^{(ij)} = \vv{M}^{(2)}$ for $i \ne j$, where
\begin{equation}
    \vv{M}^{(2)}(\vv{r}) = \frac{1}{6\pi\mu a}
    \begin{cases}
        \displaystyle \left(\frac{3a}{4r}+\frac{a^3}{2r^3}\right)\vv{I}+ \left(\frac{3a}{4r}-\frac{3a^3}{2r^3}\right)\vv{\hat{r}\hat{r}},& r>d\\
        \displaystyle \left(1-\frac{9r}{32a}\right)\vv{I}+ \frac{3r}{32a}\vv{\hat{r}\hat{r}},& r\le d
        \end{cases}
    \label{eq:rpy}
\end{equation}
and $\vv{\hat r} = \vv{r}/r$ is the unit vector from particle $j$ to particle $i$ (i.e., $\vv{r}$ = $\vv{x}_i - \vv{x}_j$). Both approximations of $\mathcal{M}$ have $\nabla_X\cdot\mathcal{M} = \mathbf{0}$, so this term can be neglected in Eq.~\eqref{eq:bd}.

We used HOOMD-blue (version 2.9.7)\cite{anderson:cms:2020} and azplugins (version 0.12.0)\cite{azplugins} to perform the BD simulations with a timestep of $\Delta t = 10^{-5} \tau_0$, where $\tau_0 = a^2/D_0$ is the natural unit of time in the BD simulations. We used the positively split Ewald technique developed by Fiore et al.~(splitting parameter 0.5, relative error tolerance $10^{-3}$) to implement the RPY hydrodynamic approximation \cite{ermak:jcp:1978, beenakker:1986, fiore:2017}. This implementation uses a periodic version of Eq.~\eqref{eq:rpy} due to the boundary conditions of the simulation box. To mitigate finite-size effects from these periodic boundary conditions, we made the edge length of the cubic simulation box twice the diameter of the droplet.

\subsubsection{Multiparticle collision dynamics}
\label{sec:methods:mpcd}
The free-draining and RPY hydrodynamic approximations used in our BD simulations neglect the influence of the droplet's liquid--vapor interface on flow. Hydrodynamic effects from spherical confinement can be incorporated in BD but add significant computational complexity \cite{aponte-rivera:2016}. Accordingly, one of our objectives was to assess the significance of hydrodynamic boundary effects on the particle distribution during drying (Sec.~\ref{sec:results:confinement}), for which we used MPCD simulations. MPCD is a mesoscale particle-based simulation method compatible with both colloidal particles and boundary surfaces\cite{malevanets:1999, howard:cpc:2018, ihle:2001, gompper:2008}. Some of us have previously shown that MPCD reproduces expected self-diffusion and sedimentation coefficients for colloidal suspensions\cite{wani:2022, wani:2024}, and we have also used it to simulate drying droplets without considering boundary effects \cite{howard:jcp:2020, yetkin:2024}. Details on our new approach for implementing the droplet boundary in MPCD are given in Sec.~\ref{sec:results:confinement}, but we summarize the other technical details of the method here.

The solvent was modeled explicitly as point particles that participated in alternating streaming and collision steps. During the streaming step, the solvent particles moved ballistically with bounce-back reflection from the droplet interface. Following this step, the solvent particles were grouped into cubic cells of side length $\ell$ and underwent a stochastic, momentum-exchanging collision with the other particles in the same cell. We used the stochastic rotation dynamics collision scheme without angular momentum conservation \cite{malevanets:1999}, where particle velocities relative to the cell's center-of-mass velocity were rotated about a randomly chosen axis by a fixed angle. The collision cells were shifted along each Cartesian direction by an amount randomly chosen from the interval $[-\ell/2,\ell/2)$ before each collision step to ensure Galilean invariance\cite{ihle:2001}. A cell-level Maxwell--Boltzmann thermostat was used to maintain constant temperature \cite{haung:2010}.

The colloidal particles were represented using a discrete-particle model \cite{wani:2022, wani:2024, poblete:2014, peng:2024}. The surface of a colloidal particle was discretized by subdividing the faces of a regular icosahedron once then scaling the resulting 42 vertices to lie on the surface of a sphere with diameter $d = 6\,\ell$. These vertex particles were bonded to their nearest neighbors and to a central particle using harmonic bonds with spring constant $5000\,k_{\rm B}T/\ell^2$. The vertex particles did not otherwise interact, and the excluded volume between colloidal particles was modeled using Eq.~\eqref{eq:wca} between central particles. The colloidal particles were coupled to the solvent by including the vertex particles in the collision step, and their positions and velocities were updated between collisions using velocity Verlet integration.

We used a solvent number density $5\,\ell^{-3}$, rotation angle $130^\circ$, and time $0.1\,\tau$ between collisions, where $\tau = \sqrt{\beta m \ell^2}$ is the natural unit of time in the MPCD simulations with $m$ being the mass of a solvent particle. The mass of the central and vertex particles was $5\,m$, and the timestep for the Verlet integration was $0.005\,\tau$. These parameters give a solvent viscosity $\mu = 3.95\,k_{\rm B}T\tau/\ell^3$, a self-diffusion coefficient $D_0 \approx 4.5 \times 10^{-3}\,\ell^2/\tau$ estimated from the Stokes--Einstein relation, and so $\tau_0 \approx 2.0 \times 10^3 \,\tau$. All MPCD simulations were performed using a modified version of HOOMD-blue 2.9.7\cite{anderson:cms:2020, howard:2018}.

\subsection{Dynamic density functional theory}
\label{sec:methods:ddft}
Our DDFT model for the drying colloidal suspension was based on Rex and L\"{o}wen's formulation that includes pairwise HIs \cite{rex:prl:2008, rex:epje:2009}. We first used their framework to establish a conservation equation for the average local particle number density $\rho(\vv{x},t)$ at position $\vv{x}$ and time $t$,
\begin{equation}
\frac{\partial \rho}{\partial t} + \nabla_\vv{x} \cdot (\vv{j}^{(1)} + \vv{j}^{(2)}) = 0,
\label{eq:ddft}
\end{equation}
where $\vv{j}^{(1)}(\vv{x},t)$ and $\vv{j}^{(2)}(\vv{x},t)$ are contributions to the particle flux, and $\nabla_\vv{x}$ denotes the vector differential operator with respect to $\vv{x}$. For simplicity, we will not explicitly denote dependence on $t$ in the rest of this discussion, but quantities that depend on the local density should be understood to vary in time.

The first flux contribution $\vv{j}^{(1)}$ is associated with the motion due to forces acting directly on the particles that is propagated by $\vv{M}^{(1)}$\cite{marconi:jcp:1999, archer_evans:2004, archer:jpcm:2005, archer:2009},
\begin{equation}
\vv{j}^{(1)}(\vv{x}) = -\rho(\vv{x}) \vv{M}^{(1)} \cdot \nabla_\vv{x} \frac{\delta A}{\delta \rho(\vv{x})},
\end{equation}
and so is the same for both the free-draining and RPY approximations used in this work. In this expression, $A$ is the Helmholtz free energy,
\begin{equation}
A[\rho] = A^{\rm ig}[\rho] + A^{\rm ex}[\rho] + \int \d{\vv{x}} \rho(\vv{x}) \psi(\vv{x}),
\end{equation}
that is a functional of $\rho$. The first term $A^{\rm ig}$ is the free energy of an ideal gas,
\begin{equation}
\beta A^{\rm ig}[\rho] = \int\d{\vv{x}} \rho(\vv{x}) \left(\ln\left[\lambda^3 \rho(\vv{x})\right]-1\right),
\end{equation}
where $\lambda$ is the thermal wavelength that accounts for integration over a particle's momentum and whose specific value will not affect our calculations; we nominally set $\lambda = d$. The second term $A^{\rm ex}$ is the excess free energy for a fluid of hard-sphere particles. The final term in $A$ is the potential energy from particles interacting with the liquid--vapor interface [Eq.~\eqref{eq:slv}].

Some of us previously tested various excess free-energy functionals for hard-sphere particles drying in a film geometry \cite{kundu:2022}, finding that Rosenfeld's fundamental measure theory \cite{Rosenfeld:1989uh, Roth:2010ei} (FMT) was more accurate and numerically stable than alternatives based on local density approximations. Hence, we have used the Rosenfeld FMT functional for $A^{\rm ex}$,
\begin{equation}
A^{\rm ex}[\rho] = \int \d{\vv{x}} \Phi(\{n_\alpha(\vv{x})\}),
\end{equation}
where $\Phi$ is a free-energy density,
\begin{align}
\beta \Phi(\{n_\alpha\}) = &-n_0 \ln(1-n_3) + \frac{n_1 n_2 - \vv{n}_1\cdot\vv{n}_2}{1-n_3} \nonumber \\
&+ \frac{n_2^3-3 n_2 (\vv{n}_2\cdot\vv{n}_2)}{24\pi(1-n_3)^2},
\end{align}
that is a function of a set of weighted densities $\{n_\alpha\} = \{n_0, n_1, n_2, n_3, \vv{n}_1, \vv{n}_2\}$. The weighted densities are functionals calculated by convolving $\rho$ with various weight functions $w_\alpha$,
\begin{equation}
n_\alpha(\vv{x}) = \int \d{\vv{y}} \rho(\vv{y}) w_\alpha(\vv{x}-\vv{y}).
\end{equation}
There are four scalar weight functions,
\begin{subequations}
\begin{align}
w_3(\vv{r}) &= \theta(a-|\vv{r}|) \\
w_2(\vv{r}) &= \delta(a-|\vv{r}|) \\
w_1(\vv{r}) &= \frac{w_2(\vv{r})}{4 \pi a} \\
w_0(\vv{r}) &= \frac{w_2(\vv{r})}{4 \pi a^2},
\end{align}
\end{subequations}
and two vector weight functions,
\begin{subequations}
\begin{align}
\vv{w}_2(\vv{r}) &= -\nabla_\vv{r} w_3(\vv{r}) = w_2(\vv{r}) \frac{\vv{r}}{|\vv{r}|} \\
\vv{w}_1(\vv{r}) &= \frac{\vv{w}_1(\vv{r})}{4 \pi a},
\end{align}
\end{subequations}
where $\theta(\vv{r})$ is the Heaviside step function and $\delta(\vv{r})$ is the Dirac delta function.

The second flux contribution $\vv{j}^{(2)}$ is associated with the motion induced by HIs between particles that is propagated by $\vv{M}^{(2)}$ \cite{rex:prl:2008,rex:epje:2009},
\begin{equation}
\vv{j}^{(2)}(\vv{x}) = - \int d\vv{y} \rho^{(2)}(\vv{x},\vv{y})\,{\vv M}^{(2)}(\vv{x}-\vv{y})\cdot\nabla_{\vv y}\frac{\delta A}{\delta\rho(\vv{y})},
\label{eq:flux_rpy}
\end{equation}
and so is zero for the free-draining approximation but nonzero for the RPY approximation. Here, $\rho^{(2)}$ is the two-body density correlation function assumed in DDFT to be that of an equivalent equilibrium system with inhomogeneous density $\rho(\vv{x})$ (the ``adiabatic'' assumption)\cite{marconi:jcp:1999, tevrugt:advphys:2020}. Although $\rho^{(2)}$ can be calculated from $A$ using a generalized Ornstein--Zernike approach \cite{hansen:2006}, it is common practice to approximate it using the radial distribution function $g$ of a homogeneous system for computational convenience \cite{rex:prl:2008, rex:epje:2009, goddard:2013, donev:2014, goddard:2016, goddard:2020},
\begin{equation}
\rho^{(2)}(\vv{x},\vv{y}) \approx \rho(\vv{x})\rho(\vv y)g(|\vv{y}-\vv{x}|; \bar{\rho})
\label{eq:twobody}
\end{equation}
where $g$ is evaluated at the mean of the local densities at $\vv{x}$ and $\vv{y}$, $\bar\rho = [\rho(\vv{x}) + \rho(\vv{y})]/2$. We will consider both dilute and nondilute approximations of $g$ in this work.

In contrast to our particle-based simulations, DDFT can leverage symmetry to reduce model dimensionality. Specifically, we expected that density gradients should only form in the droplet in the radial direction and so that the average local particle density $\rho(x)$ should only be a function of the radial distance $x = |\vv{x}|$ from the center of the droplet. We used this radial symmetry to simplify the calculation of the weighted densities (Appendix~\ref{sec:appendix_A}). When $x \ge a$, the scalar weighted densities take the form
\begin{equation}
n_\alpha(x) = \frac{1}{x} \int \d{y} y \rho(y) w_\alpha(x-y)
\label{eq:nsphere}
\end{equation}
with
\begin{subequations}
\label{eq:wsphere}
\begin{align}
w_3(r) &= \pi(a^2-r^2)\theta(a-|r|) \label{eq:wsphere:w3}\\
w_2(r) &= 2 \pi a \theta(a-|r|) \label{eq:wsphere:w2}
\end{align}
\end{subequations}
and $w_1$ and $w_0$ following by proportionality to $w_2$. For the vector weighted densities,
\begin{align}
\vv{n}_2(x) = \frac{1}{x} \bigg[n_3(x)
+ \int \d{y} y \rho(y) 2 \pi (x-y) \theta(a-|x-y|) \bigg] \vv{\hat{x}}
\label{eq:n2sphere_outer}
\end{align}
where $\vv{\hat{x}}$ is the radial basis vector and $\vv{n}_1$ follows by proportionality. These expressions are convolutions and so can be efficiently evaluated using a numerical Fourier transform of $x \rho(x)$ and analytical Fourier transforms of the effective one-dimensional weight functions\cite{sear:jcp:2003}. The case where $x < a$ must be treated differently, giving
\begin{subequations}
\label{eq:ninner}
\begin{align}
n_3(x) &= 4 \pi \int_0^{a-x} \d{y} \rho(y) y^2 \nonumber \\
&+ \frac{\pi}{x} \int_{a-x}^{x+a} \d{y} \rho(y) y \left[a^2 - (x-y)^2 \right], \label{eq:ninner:n3} \\
n_2(x) &= \frac{2 \pi a}{x} \int_{a-x}^{x+a} \d{y} \rho(y) y, \label{eq:ninner:n2} \\
\vv{n}_2(x) &= \left[\frac{\pi}{x^2} \int_{a-x}^{x+a} \d{y} \rho(y) y \left(a^2 + x^2 - y^2 \right)\right] \vv{\hat{x}}, \label{eq:ninner:nv2}
\end{align}
\end{subequations}
and the rest again by proportionality. These expressions are not convolutions and so the integrals must be evaluated directly using quadrature. Because of the radial symmetry of $\rho$, we also expected the fluxes $\vv{j}^{(1)}(x)$ and $\vv{j}^{(2)}(x)$ to be nonzero only in the radial direction. The evaluation of $\vv{j}^{(1)}$ under this symmetry is straightforward, but we will discuss the simplification of $\vv{j}^{(2)}$ in detail in Sec.~\ref{sec:results:ddftrpy}.

To implement our DDFT model numerically, we represented $\rho(x)$ on a regular mesh with spacing $0.01\,d$, with the local density defined at the center of each mesh volume. Before beginning drying, $\rho(x)$ was initialized uniformly inside the droplet at a value consistent with $\eta_0$. We then used fixed-point iteration to solve for the equilibrium local density profile that minimizes the grand potential when the average number of particles is constrained to be $N$. Drying was simulated by evolving $\rho$ according to Eq.~\eqref{eq:ddft} using a finite-volume scheme and the implicit Euler method for time integration (timestep $10^{-5}\,d^2/D_0 = 4 \times 10^{-5}\,\tau_0)$. Most gradients were evaluated directly using finite-difference approximations, but the gradient of $\delta A^{\rm ig}/\delta \rho$ was first simplified analytically. Most integrals were evaluated using the midpoint rule, but Eq.~\eqref{eq:ninner} was evaluated using the trapezoidal rule with bin width $0.005\,d$ to improve accuracy. Linear interpolation was used to obtain $\rho$ at points between mesh centers.

\section{Results and discussion}
\label{sec:results}
This section is divided into four parts, with a summary of the content and main findings given here. In Sec.~\ref{sec:results:fd}, we evaluate the accuracy of the DDFT model for drying colloidal suspensions in spherical confinement by comparing it to reference BD simulations assuming free-draining HIs, finding excellent agreement. In Sec.~\ref{sec:results:hi}, we analyze the effects of including unconfined pairwise HIs on the distribution of colloidal particles during drying by performing BD simulations with pairwise HIs (BD+RPY), finding that larger particle density gradients form when pairwise HIs are present. In Sec.~\ref{sec:results:confinement}, we examine how the boundary conditions on the solvent modify the distribution of colloidal particles during drying by comparing MPCD simulations with both unconfined and confined solvent, finding little difference. Finally, in Sec.~\ref{sec:results:ddftrpy}, we implement pairwise HIs in our DDFT model (DDFT+RPY), assess its accuracy against BD+RPY simulations, and identify limitations of some approximations made in constructing the DDFT+RPY model that should be carefully considered.

\subsection{DDFT with free-draining HIs}
\label{sec:results:fd}
To establish a baseline expectation of accuracy for the DDFT+RPY model (Sec.~\ref{sec:results:ddftrpy}), we first compared the DDFT model to BD simulations assuming free-draining HIs in both. For a given initial volume fraction $\eta_0$, we computed the local density profile $\rho(x)$ when 0\%, 25\%, 50\%, 75\% and 100\% of the required drying time had passed. Equilibrium local density profiles were also obtained at the equivalent average $\eta$ for each of these time points. The local density profiles for the BD simulations were computed using histograms with bin width $0.25\,d$ averaged over four independent simulations. We then numerically calculated the volume fraction profile $\eta(x)$, which is equivalent to $n_3(x)$, using the numerical approach described in Sec.~\ref{sec:methods:ddft}. For the BD simulations, we do not show $\eta(x)$ for $x < 5\,d$ due to poor statistics in this region.

\begin{figure*}
    \centering
    \includegraphics{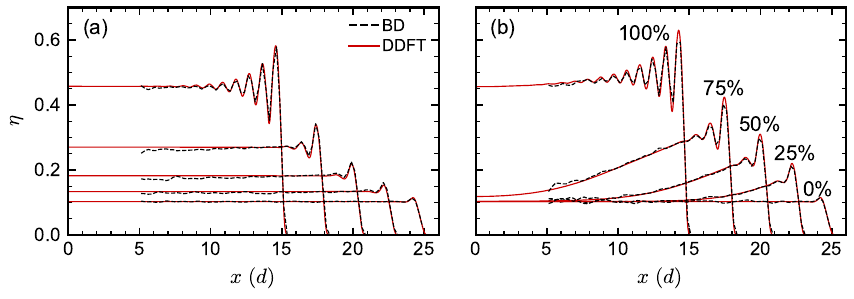}
    \caption{Volume fraction $\eta$ at radial distance $x$ from center of droplet obtained from BD simulations and DDFT with free-draining HIs under (a) equilibrium conditions and (b) drying conditions with ${\rm Pe} = 10$ for $\eta_0 = 0.10$. Profiles in (b) are shown at 0\%, 25\%, 50\%, 75\% and 100\% of the total drying time required to reach a final average droplet volume fraction of 0.50, and the equilibrium profiles in (a) were computed for the corresponding average droplet volume fraction.}
    \label{fig:bd_vs_ddft_free_draining}
\end{figure*}

The equilibrium volume fraction profiles [Fig.~\ref{fig:bd_vs_ddft_free_draining}(a)] showed excellent agreement between BD and DDFT, as expected for the Rosenfeld FMT functional \cite{kierlik:1991, Roth:2010ei}. The equilibrium volume fraction profiles were inhomogeneous near the droplet's liquid--vapor interface, exhibiting oscillations with approximate period $d$. Both the amplitude of the oscillations and the distance from the interface over which they persisted increased with the average volume fraction (or decreasing $R$), a characteristic of hard spheres in confinement\cite{snook:1978}. The volume fraction profiles during drying [Fig.~\ref{fig:bd_vs_ddft_free_draining}(b)] were also in nearly quantitative agreement between BD and DDFT, which is consistent with results from prior work for thin films \cite{howard:lng:2017, kundu:2022}. Due to the advection-dominated drying (${\rm Pe} = 10$), substantial gradients in $\eta(x)$ developed as drying progressed, with this gradient propagating toward the droplet center over time. The volume fraction near the droplet's liquid--vapor interface and corresponding oscillations in $\eta(x)$ tended to be larger during drying than in equilibrium at equivalent average volume fractions (e.g., at 50\% or 75\% of the total drying time). However, at the end of drying, $\eta(x)$ had a smaller gradient and more closely resembled the equivalent equilibrium profile compared to earlier stages. This behavior reflects the tendency of particles to concentrate more uniformly in the droplet after the gradient propagates to its center. Figure~\ref{fig:bd_vs_ddft_free_draining} shows the results for only $\eta_0 = 0.10$, but similar behavior was obtained for other values of $\eta_0$ (Fig.~S1).

\subsection{Effect of HIs between particles}
\label{sec:results:hi}
\begin{figure*}
    \centering
    \includegraphics{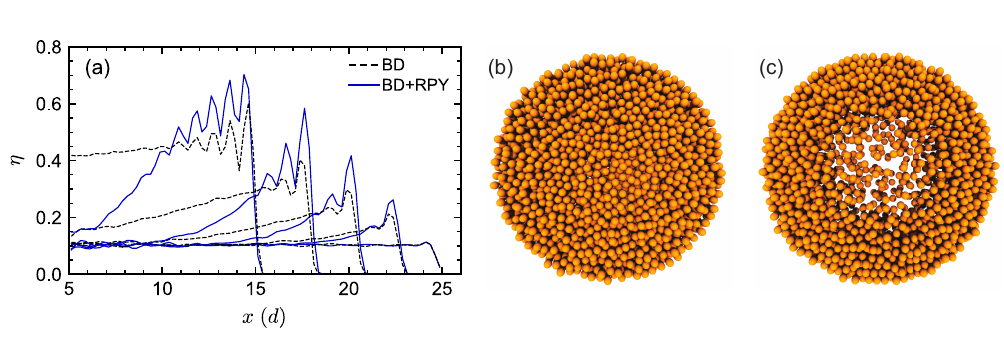}
    \caption{(a) Volume fraction $\eta$ at radial distance $x$ from center of droplet obtained using BD and BD+RPY simulations under the same conditions as in Fig.~\ref{fig:bd_vs_ddft_free_draining}(b). 
    (b,c) Cross-sectional image of particles in the final structures from the (b) BD and (c) BD+RPY simulations. The cross section has thickness $10d$ and is centered on the droplet, and the images were rendered using VMD 1.9.4.\cite{vmd}}
    \label{fig:bd_vs_bd_rpy}
\end{figure*}

After simulating drying with free-draining HIs, we next explored the impact of including pairwise HIs between particles on their distribution within the drying droplet using particle-based simulations. We repeated our BD simulations using the RPY mobility tensor as described in Sec.~\ref{sec:methods:bd}; we will refer to these as BD+RPY simulations. In Fig.~\ref{fig:bd_vs_bd_rpy}(a), we plot the volume fraction profiles from the BD and BD+RPY simulations at the same time points, revealing an enhancement of particles near the droplet interface and corresponding smaller concentration near the droplet center in the BD+RPY simulations. This difference is also evident in cross-sectional images taken at the end of the BD and BD+RPY simulations [Figs.~\ref{fig:bd_vs_bd_rpy}(b) and \ref{fig:bd_vs_bd_rpy}(c), respectively]. Similar trends were observed for the other $\eta_0$ values (Fig.~S2) as well as in prior simulations of a drying film with and without HIs \cite{howard:jcp:2018}, where the enhanced particle concentration at the film interface promoted crystallization.

We attribute the pronounced difference in particle distributions in the BD and BD+RPY simulations to two physical effects caused by the presence of pairwise HIs, as outlined by Batchelor for a dilute sedimenting suspension \cite{batchelor:jfm:1972}. First, colloidal particle suspensions can be reasonably assumed to be incompressible, such that a flux of particles implies an opposing flow of solvent in order to conserve volume. Hence, the inward diffusion of particles (toward the center of the droplet) should be accompanied by an outward ``backflow'' of solvent (toward the droplet's liquid--vapor interface). This backflow will tend to impede particle motion, so a larger particle concentration will develop near the droplet's liquid--vapor interface with backflow than if the solvent were stationary. Second, pairwise HIs enhance this effect because solvent moving with a diffusing particle in a volume that is excluded to other particles must be balanced by backflow that is experienced elsewhere. The BD+RPY simulations assume an incompressible suspension and include pairwise HIs, whereas the BD simulations neglect the hydrodynamic presence of other particles and there is no backflow or pairwise HIs (i.e., the solvent is effectively stationary).

To more quantitatively understand these effects, we can consider the sedimentation coefficient $K$ that relates the mean particle sedimentation velocity to the sedimentation force \cite{batchelor:jfm:1972}. In BD simulations of bulk sedimentation, $K = 1$. Accounting for the backflow of solvent required to maintain incompressibility (but otherwise neglecting the presence of HIs) gives $K = 1 - \eta$, i.e., the backflow hinders sedimentation. Finally, the inclusion of pairwise HIs through the RPY tensor gives $K \approx 1 - 5\eta$ under dilute conditions \cite{wani:2022, ladd:1990, brady:pf:1988, banchio:2008}, i.e., sedimentation is more hindered when pairwise HIs are included than when they are not. We hence expect larger concentration gradients to develop in the BD+RPY simulations than in the BD simulations, consistent with the results shown in Fig.~\ref{fig:bd_vs_bd_rpy}.

\subsection{Effect of solvent boundary conditions}
\label{sec:results:confinement}
Although the BD+RPY simulations presented in Sec.~\ref{sec:results:hi} provide a more realistic description of HIs than free-draining BD simulations, they still do not consider the influence of the droplet's liquid--vapor interface on the HIs. The presence of this interface requires changes to $\vv{M}^{(1)}$ and $\vv{M}^{(2)}$\cite{aponte-rivera:2016}, potentially affecting both the individual and collective motion of the particles and hence also their distribution in the droplet during drying. However, these changes are mathematically involved, and including them significantly complicates the numerical implementation of BD+RPY simulations and DDFT calculations. Hence, here we used MPCD simulations (Sec.~\ref{sec:methods:mpcd}), for which it is more convenient to incorporate boundaries \cite{whitmer:2010}, to test for the significance of boundary effects.

\begin{figure}
    \centering
    \includegraphics{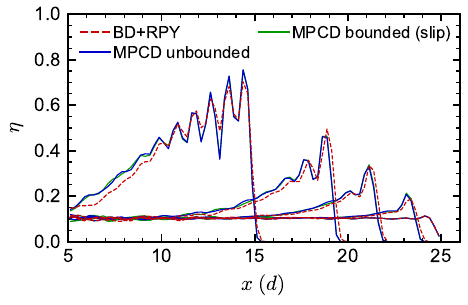}
    \caption{Volume fraction $\eta$ at radial distance $x$ from the droplet center obtained from a BD+RPY simulation, an MPCD simulation with an unbounded solvent, and an MPCD simulation with a slip boundary condition on the solvent at the droplet's liquid--vapor interface for the same conditions as in Fig.~\ref{fig:bd_vs_ddft_free_draining}(b). Note that the results from the two MPCD simulations are essentially overlapping.}
    \label{fig:bd_rpy_vs_mpcd}
\end{figure}

We first tested whether MPCD produced the same results as our BD+RPY simulations when boundary effects were neglected. We allowed the MPCD solvent to freely flow through the nominal droplet interface and hence fill the entire simulation box with solvent particles. The P{\'e}clet number in the MPCD simulations was matched to that in the BD+RPY simulations using the Stokes--Einstein estimate of the particle self-diffusion coefficient in the MPCD solvent at infinite dilution. The volume fraction profiles of the colloidal particles computed from the MPCD simulations were in very good agreement with the same from the BD+RPY simulations throughout the entire simulation (Fig.~\ref{fig:bd_rpy_vs_mpcd}); however, we did observe some small quantitative discrepancies, which could be due to several reasons. First, the MPCD simulations do not assume pairwise RPY HIs. As a result, MPCD and BD+RPY simulations of colloidal suspensions can have different concentration-dependent self-diffusion and sedimentation coefficients; some of us have previously shown that the MPCD simulations are in somewhat closer agreement with experimental data \cite{wani:2022}. Second, the P\'{e}clet number in the MPCD simulations might be imperfectly matched to the BD+RPY simulations because the true self-diffusion coefficient of a particle in the MPCD solvent can differ from the Stokes--Einstein estimate by a few percent \cite{wani:2022, wani:2024, poblete:2014, peng:2024}. Finally, we observed that a small solvent density gradient developed in the MPCD simulations, with the solvent density being larger than average near the droplet center and less than average near the interface (Fig.~S3). The BD+RPY simulations assume an incompressible solvent, which we regard as more physically realistic.

We attribute the solvent density gradient in the MPCD simulations to the compressibility of the MPCD solvent \cite{zantop:2021}. Pressure gradients in the suspension are expected to relax more quickly (by flow) than concentration gradients (by diffusion). Hence, an osmotic pressure gradient caused by an inhomogeneous distribution of particles must be balanced by a solvent pressure gradient to maintain constant total pressure \cite{sear:pre:2017}; if the solvent is compressible, this pressure gradient leads to a corresponding solvent density gradient. To support this hypothesis, we ran an additional MPCD simulation at ${\rm Pe} = 100$, which should produce a stronger osmotic pressure gradient than at ${\rm Pe} = 10$, and we indeed found that the solvent density gradient became more pronounced (Fig.~S3). Because the properties of the MPCD solvent depend on its number density, a solvent density gradient also produces a viscosity gradient that may affect the distribution of particles in the drying droplet. We note, however, that the observed variation in solvent density was only 2.5\% of its average for ${\rm Pe} = 10$ (Fig.~S3), which translates to a variation of 3.0\% in the viscosity. We hence believe that this effect does not significantly alter the results presented here. Nevertheless, we recommend caution when simulating at larger P\'{e}clet numbers that may produce larger colloidal particle concentration gradients and hence larger solvent density gradients.

\begin{figure}
    \centering
    \includegraphics{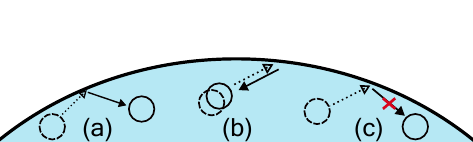}
    \caption{Schematic of different bounce-back conditions for solvent particles in MPCD simulations: (a) slip, (b) no-slip, and (c) deletion.}
    \label{fig:mpcd}
\end{figure}

Given the overall good agreement between the MPCD and BD+RPY simulations for the unbounded solvent, we extended our MPCD simulations to include either a slip or no-slip boundary condition on the solvent at the droplet interface. In our simulations, we first determined whether a solvent particle had collided with the droplet interface by checking if it was outside the droplet at the end of the streaming step. If a solvent particle was outside, we returned it to the point where it crossed the interface by calculating the amount of time $\Delta t_{\rm s}$ that it was outside the droplet (and hence remained for streaming). To perform this calculation, we approximated the radial velocity of the interface by its average value during the full streaming step $V$ ($< 0$), which allowed us to determine $\Delta t_{\rm s}$ as the smaller root of a quadratic equation,
\begin{align}
    \Delta t_{\rm s} = \frac{1}{|\vv{v}|^2-V^2} &\Bigg(\vv{x} \cdot \vv{v} - RV - \Big[(\vv{x} \cdot \vv{v} - RV)^2 \nonumber \\
    &- (|\vv{v}|^2-V^2) (|\vv{x}|^2-R^2)\Big]^{1/2}\Bigg),
\end{align}
where $\vv{x}$ is the position of the particle at the end of streaming and $R$ is the radius of the droplet at the end of streaming ($|\vv{x}| > R$). The case that $|\vv{v}|^2 = V^2$ must be evaluated as a special limit,
\begin{equation}
    \Delta t_{\rm s} = \frac{|\vv{x}|^2 - R^2}{2(\vv{x} \cdot \vv {v} - RV)},
\end{equation}
which in practice we only used when the two differed by less than $10^{-8}\,\ell^2/\tau^2$. After being returned to the interface, the solvent particle's velocity relative to the interface was reflected according to the standard bounce-back schemes \cite{lamura:2001, whitmer:2010} for a slip [only normal component, Fig.~\ref{fig:mpcd}(a)] or no-slip [both normal and tangential components, Fig.~\ref{fig:mpcd}(b)] boundary. Streaming was then continued with the reflected velocity for the remaining time $\Delta t_{\rm s}$, and the process repeated until the full streaming step was complete.

Because all solvent particles now remained in the droplet, we needed to remove some solvent particles to maintain a constant solvent density---and thus consistent solvent properties---during drying. To this end, we calculated the expected number of solvent particles in the droplet at the end of the streaming step and randomly removed a subset of the particles that had bounced back from the droplet interface during that step [Fig.~\ref{fig:mpcd}(c)]. Sometimes, the number of solvent particles requiring removal exceeded the number that bounced back, resulting in a small excess of solvent particles in the droplet; however, this excess was typically corrected for on subsequent streaming steps, with some exceptions at late stages of drying. Following best practice for MPCD simulations, we also introduced ``virtual'' solvent particles outside the droplet boundary that participated in the collision step \cite{lamura:2001, whitmer:2010, bolintineanu:2012} to prevent underfilling of the collision cells that were sliced by the droplet boundary.

We tested our approach by first simulating a drying droplet containing only pure solvent, finding that the solvent density was homogeneous and there was no outward flow, as expected theoretically (Fig.~S4). We then simulated a particle suspension in the drying droplet using slip boundary conditions for the solvent at the droplet interface. We found that the resulting particle distribution in the droplet during drying was virtually indistinguishable from that when the solvent was unbounded by the interface (Fig.~\ref{fig:bd_rpy_vs_mpcd}). We repeated the simulations using no-slip boundary conditions for the solvent particles and similarly found no notable differences in the particle distribution (Fig.~S5). From these results, we concluded that solvent boundary effects on the particle distribution could be reasonably neglected under the conditions simulated.

\subsection{DDFT with pairwise HIs}
\label{sec:results:ddftrpy}
Building on our findings that DDFT accurately captures the particle distribution in a drying droplet for free-draining HIs (Sec.~\ref{sec:results:fd}) and that particle-based simulations show pairwise HIs influence this distribution (Sec.~\ref{sec:results:hi}) but boundary effects on these HIs can be reasonably neglected (Sec.~\ref{sec:results:confinement}), we proceeded to develop a DDFT model incorporating pairwise HIs using the unconfined RPY mobility tensor. As discussed in Sec.~\ref{sec:methods:ddft}, we can leverage radial symmetry to simplify the equations that must be solved numerically. We applied this symmetry to evaluate the free-energy functional in Sec.~\ref{sec:results:fd} and now extend it to the hydrodynamic contribution to the flux, $\vv{j}^{(2)}$.

Given that the local particle density varies only in the radial direction, we expect that the effective force $\vv{f}(\vv{y}) = -\nabla_\vv{y} \delta A / \delta\rho(\vv{y})$ in Eq.~\eqref{eq:flux_rpy} should have only a radial component that depends on the radial distance, $\vv{f}(\vv{y}) = f(y) \vv{\hat{y}}$, where $y = |\vv{y}|$ and $\vv{\hat{y}} = \vv{y}/y$. The flux should similarly be nonzero only in the radial direction and be only a function of radial distance, $\vv{j}^{(2)}(\vv{x}) = j^{(2)}(x) \vv{\hat{x}}$ with $\vv{\hat{x}} = \vv{x} / x$. Using this symmetry and two-body density correlations approximated by Eq.~\eqref{eq:twobody}, it can be shown that
\begin{equation}
j^{(2)}(x) = \int \d{y} \rho(x) \rho(y) M^{(2)}(x, y; \bar \rho) f(y),
\label{eq:flux_rpy_sym}
\end{equation}
where the effective mobility $M^{(2)}$,
\begin{equation}
M^{(2)}(x,y;\bar\rho) = 2\pi y^2 \int \d{\cos\phi} \, g(r; \bar \rho) (\vv{\hat{x}} \cdot \vv{M}^{(2)} \cdot \vv{\hat{y}}),
\label{eq:effm2}
\end{equation}
is obtained by projecting $\vv{M}^{(2)}$ onto $\vv{\hat{x}}$ and $\vv{\hat{y}}$, and subsequently integrating over the angle $\phi$ between $\vv{x}$ and $\vv{y}$, with $r = (x^2 + y^2 - 2xy \cos\phi)^{1/2}$ being the distance between $\vv{x}$ and $\vv{y}$. Since $g$ always prohibits overlap for hard particles, only the projection of $\vv{M}^{(2)}$ for the case where $r > d$ is required,
\begin{align}
\vv{\hat{x}} &\cdot \vv{M}^{(2)} \cdot \vv{\hat{y}} = \frac{1}{8\pi\mu r} \Bigg[\left(1 + \frac{2 a^2}{3 r^2}\right)\cos\phi \nonumber \\
&+ \left(1 - \frac{2a^2}{r^2}\right)\left(\cos\phi - \frac{xy(1-\cos^2\phi)}{r^2}\right) \Bigg].
\end{align}
The total flux is then $\vv{j}(\vv{x}) = j(x) \vv{\hat{x}}$ with $j = j^{(1)} + j^{(2)}$, where $j^{(1)}(x) = \rho(x) f(x) / (6 \pi \mu a)$ follows straightforwardly from $\vv{j}^{(1)}$.

We initially assumed that the bulk pair correlation function could be approximated as that of a dilute hard-sphere suspension, $g(r; \bar \rho) \approx \theta(r - d)$. We emphasize that we do not regard this description to be a realistic representation of the pair correlations for all volume fractions we are considering (Fig.~S9). However, we have employed it as a first-order approximation of the true pair correlations because it has been used previously in DDFT \cite{rex:prl:2008, rex:epje:2009, goddard:2013, goddard:2016, goddard:2020} and also permits analytical simplification of Eq.~\eqref{eq:effm2}. We will discuss the accuracy and implications of this approximation below. 

To evaluate Eq.~\eqref{eq:effm2} using the dilute hard-sphere approximation of $g$, the limits of integration for $\cos\phi$ must be restricted such that two spheres do not overlap (Fig.~S6). The shortest distance between the two spheres is $|x - y|$ (when $\cos \phi = 1$). Hence, when $|x-y| > d$, there is never any overlap and $-1 \le \cos\phi \le 1$. However, when $|x-y| \le d$, the spheres can come into contact for small values of $\phi$, and the integrand is nonzero only when  $-1 \le \cos\phi \le (x^2+y^2-d^2)/(2xy)$. Evaluating Eq.~\eqref{eq:effm2} with these bounds, for which $g = 1$, we find that $M^{(2)}$ is nonzero only in the latter case, meaning
\begin{align}
M^{(2)}&(x,y) = \frac{5}{96 \mu d x^2} (d-x-y)(d+x-y) \nonumber \\
&\times (d-x+y)(d+x+y) \theta(d-|x-y|).
\label{eq:m2effdilute}
\end{align}
We note that there is an additional restriction on the limits of integration for $y$ in Eq.~\eqref{eq:flux_rpy_sym} in the special case when $x < d$. Because the longest distance between the two spheres is $x + y$ (when $\cos\phi = -1$), we must additionally require $y \ge d-x$ to prevent overlap when $x < d$.

Before performing numerical calculations with Eq.~\eqref{eq:m2effdilute}, we first considered a hypothetical case to understand how the spherical geometry of the droplet may influence particle motion compared to bulk. Analogous to bulk sedimentation \cite{batchelor:jfm:1972}, we assume that the particle density $\rho$ is uniform and the force $f$ is constant. With these assumptions, the flux $j^{(2)}$ is an integral over $M^{(2)}$ that can be taken analytically. We then define the sedimentation coefficient from the total flux in the usual way, $j = K \rho f / (6 \pi\mu a)$, giving
\begin{align}
K = \begin{cases}
    \displaystyle 1 - 5 \eta \left(1 - \frac{d^2}{5 x^2}\right),& x > d \\
    \displaystyle 1 - 5 \eta \left(\frac{x}{d} - \frac{x^3}{5 d^3}\right), & 0 < x \le d
\end{cases}.
\label{eq:Ksphere}
\end{align}
The case where $x = 0$ is intentionally excluded because the flux must be zero at the droplet center by symmetry. Focusing on the case $x > d$, we observe that $K$ in the spherical geometry is similar to that of a bulk suspension,\cite{batchelor:jfm:1972} and the hydrodynamic flux $j^{(2)}$ tends to inhibit motion of the particles. However, there is an additional term that reduces this inhibition and is qualitatively associated with the curvature of the spherical geometry. Indeed, the limiting value of $K$ as $x \to \infty$, where curvature effects should become negligible, is the same as for a bulk suspension of dilute hard spheres with RPY HIs\cite{wani:2022, ladd:1990, brady:pf:1988, banchio:2008}. In practice, the contribution to $K$ from curvature is less than 1\% when $x = 5\,d$. Overall, the behavior in this hypothetical case is qualitatively consistent with our simulations showing that HIs inhibited particle diffusion during drying [Fig.~\ref{fig:bd_vs_bd_rpy}(a)].

\begin{figure}
    \centering
    \includegraphics{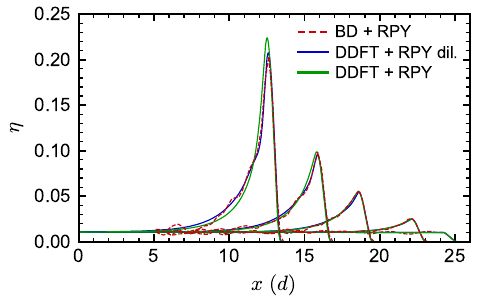}
    \caption{Volume fraction $\eta$ at radial distance $x$ from center of droplet obtained from BD+RPY simulations, as well as DDFT calculations with RPY HIs assuming either dilute or concentration-dependent pair correlations. The initial volume fraction was $\eta_0 = 0.01$, and the P\'{e}clet number was ${\rm Pe} = 10$. The curves are shown at approximately $0\%$, $25\%$, $50\%$, $75\%$, and $92\%$ of the total drying time required to reach a final average droplet volume fraction of 0.10.} 
    \label{fig:bd_rpy_vs_ddft_rpy}
\end{figure}

We next performed DDFT calculations for particles in a drying droplet starting from a small initial volume fraction of $\eta_0 = 0.01$ and ending at a final volume fraction $\eta = 0.1$ to be consistent with our dilute approximation of $g$. We also conducted BD+RPY simulations starting from the same volume fraction to establish a reference. We found that the DDFT calculations were in excellent agreement with the BD+RPY simulations during the early stages of drying (Fig.~\ref{fig:bd_rpy_vs_ddft_rpy}). However, the DDFT calculations became unphysical after $97\,\%$ of the total drying time, when the average volume fraction in the droplet was approximately 0.085 and the maximum local volume fraction was 0.29. At this point, particles were ejected from the droplet in the DDFT calculations, which is wholly inconsistent with the BD+RPY simulations. We repeated our DDFT calculations starting from $\eta_0 = 0.10$, the same as in Fig.~\ref{fig:bd_vs_ddft_free_draining}, and we again observed the same effect occurring after $30\,\%$ of the total drying time, when the average volume fraction was 0.14 and the maximum local volume fraction was 0.39. 

We hypothesized that this unphysical behavior in the DDFT calculations, despite initially good agreement with the BD+RPY simulations, might be due to the dilute approximation of the pair correlation function $g$. This hypothesis is partially supported by our analysis of $K$ in the case of sedimentation. In Eq.~\eqref{eq:Ksphere}, $K$ can become negative when $\eta > 0.2$, meaning that particles actually move opposite the applied force, while in reality, we expect that $K$ should become small but remain positive for a concentrated suspension\cite{russel:1989}. This limitation of the dilute approximation of $g$ has been noted elsewhere for sedimentation of bulk suspensions \cite{brady:pf:1988}. We note, though, that there is not a direct translation from Eq.~\eqref{eq:Ksphere} to our drying calculations because the particle distribution in the droplet is inhomogeneous.

To attempt to circumvent the challenges we encountered using the dilute approximation of $g$, we used a more accurate, density-dependent expression for $g(r;\bar\rho)$. We adopted the analytical form of $g$ developed by Trokhymchuk et al., which is highly accurate for bulk hard-sphere suspensions across a range of volume fractions (Fig.~S7) \cite{trokhymchuk:2005}. However, this form of $g$ is not readily amenable to analytical integration. Therefore, we constructed a numerical model for $M^{(2)}(x, y; \bar\rho)$ for use in our DDFT calculations based on three-dimensional linear interpolation. To obtain a more accurate interpolation, we restricted $0\,d \le x \le 5\,d$ because we found $M^{(2)}$ did not vary significantly for larger values of $x$, clamping larger values of $x$ to the upper bound of this domain. We also reexpressed $y = x + \Delta x$ as a variation $\Delta x$ about $x$ and restricted $-3\,d \le \Delta x \le 3\,d$, assuming contributions at larger separations were negligible. We then used linear interpolation over this domain of $x$ and $\Delta x$ with $\bar\rho$ varying over the domain consistent with $0 \le \eta \le 0.60$, using a spacing of 0.05 for each. More information regarding the construction of this interpolation scheme can be found in Sec.~\ S3. We verified that our approach produced similar volume fraction profiles as our previous DDFT calculations for dilute conditions (Fig.~\ref{fig:bd_rpy_vs_ddft_rpy}), albeit somewhat less accurate as the particles became more concentrated. However, we ultimately found that the DDFT calculations still failed at a similar point as when we used the dilute approximation of $g$. Evidently, improving the accuracy of $g(r;\bar \rho)$ on its own was insufficient to improve the DDFT calculation.

To better understand why our DDFT calculations were still failing, we conducted equilibrium Monte Carlo simulations of hard-sphere particles to measure their true two-body density correlation function when they are distributed inhomogeneously inside a droplet. We considered five different volume fractions $\eta = 0.10, 0.20, 0.30, 0.40$, and $0.50$ inside a droplet with radius $25\,d$. We first equilibrated configurations at different volume fractions using BD. Particles were initially placed randomly inside the droplet. During this step, the particles interacted with each other via Eq.~\eqref{eq:wca} using a larger nominal particle diameter of $1.05\,d$ and with the interface via Eq.~\eqref{eq:interface} using a smaller nominal droplet radius of $24\,d$, which ensured the absence of overlaps between particles or between particles and the interface when we later switched to true hard-sphere interactions. These simulations were carried out for $10^4 \,\tau_0$, and configurations were saved every $100 \, \tau_0$. Starting from these configurations, we then ran 100 independent hard-sphere Monte Carlo simulations using the true particle diameter $d$ and droplet radius $25\,d$. For computational convenience, the droplet interface was now also represented by a hard potential rather than a harmonic potential. We used HOOMD-blue (version 4.7.0)\cite{anderson:cpc:2016} with an initial maximum particle displacement of $0.2\,d$, which we subsequently tuned every 100 steps during the first 5000 steps of the simulation so that the move acceptance probability was approximately $0.5$. For our simulation configuration, a step is defined as attempting 2 trial moves per particle. We ran each simulation for $5.1 \times 10^5$ steps and saved the final configuration from each simulation for analysis.

Using the radial symmetry of the droplet, we then calculated $\rho^{(2)}$ as a function of the distance of the reference particle from the center of the droplet and the displacement vector between the two particles. Specifically, for a reference particle at $\vv{x}$ and second particle at $\vv{y}$, we calculated $\rho^{(2)}(x, r, \cos\phi_r)$ where $x = |\vv{x}|$, $r = |\vv{y}-\vv{x}|$, and $\cos\phi_r = [(\vv{y}-\vv{x}) \cdot \vv{x}] / (r x)$ using a multidimensional histogram with bin sizes $0.1\,d$ for $5\,d \le x \le 25\,d $, $0.1\,d$ for $1\,d \le r \le 5\,d$, and 0.1 for $-1 \le \cos \phi_r \le 1$. The two-body density correlations measured from the simulations in this way will be referred to as $\rho^{(2)}_{\rm s}$. We then evaluated our approximation of the two-body density correlation function [Eq.~\eqref{eq:twobody}] at the same coordinates as $\rho^{(2)}_{\rm s}$ using the density-dependent form of $g(r; \bar\rho)$ from Ref.~\citenum{trokhymchuk:2005}. For this purpose, the average local density $\rho(x)$ was measured from the Monte Carlo simulations (histogram bin size $0.1\,d$) and used to evaluate $\bar\rho$. The mean density $\bar\rho$ input to $\rho^{(2)}_{\rm a}$ was clipped to $\bar \rho \le 1.15\,d^{-3}$, which corresponds to a volume fraction of 0.60. The two-body density correlations calculated in this way will be referred to as $\rho^{(2)}_{\rm a}$.

\begin{figure*}
    \centering
    \includegraphics{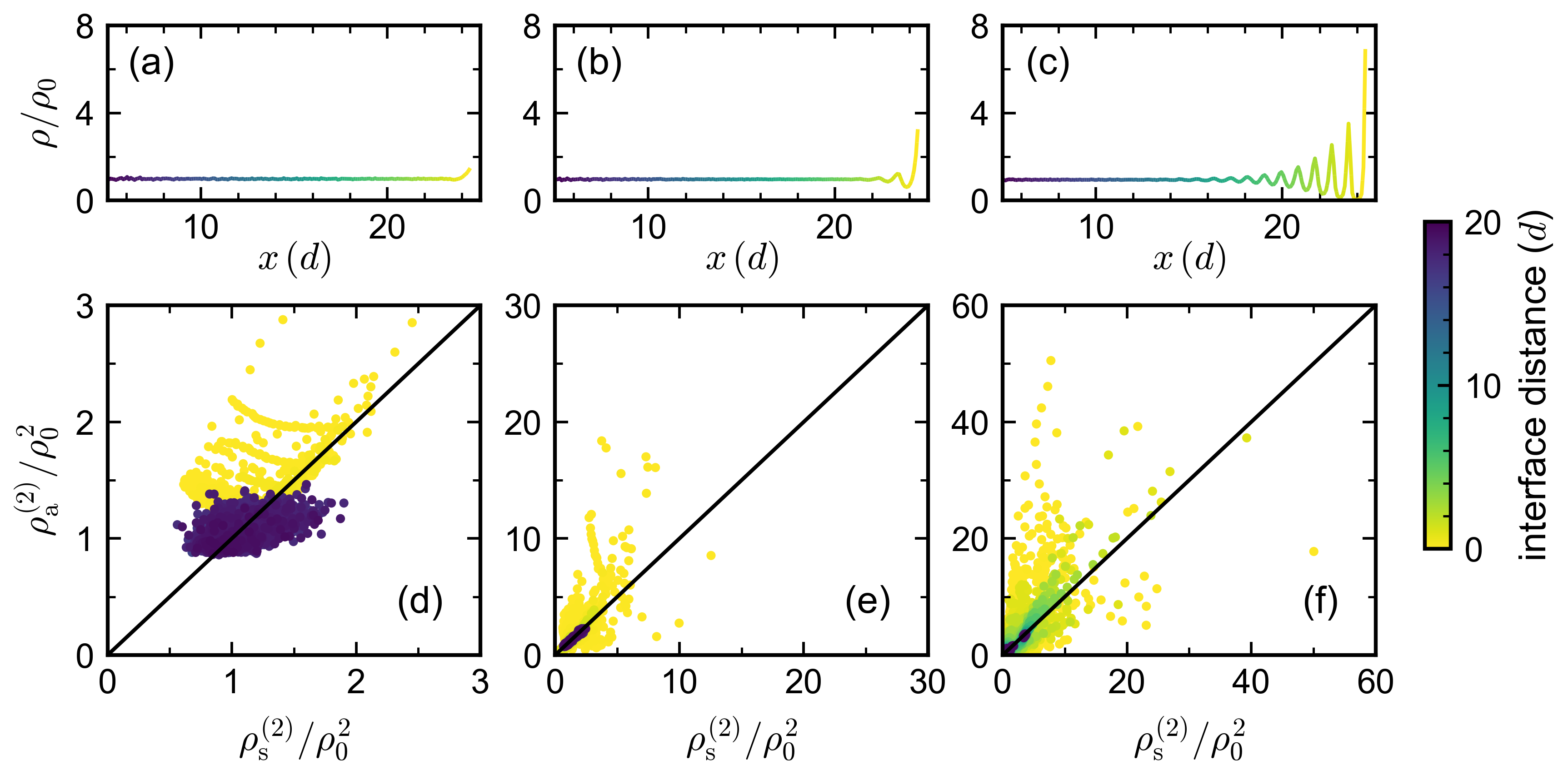}
    \caption{(a--c) Local density $\rho$ at radial distance $x$ from center of droplet for hard-sphere particles from Monte Carlo simulations at average volume fraction (a) 0.10, (b) 0.30, and (c) 0.50. The local density is normalized by the average density of point particles in the droplet accounting for the hard interaction with the droplet interface, $\rho_0 = 3 N / [4 \pi (R-a)^3]$. (d--f) Approximated two-body density correlation $\rho^{(2)}_{\rm a}$, using Eqs.~\eqref{eq:twobody} with $g$ given by Ref.~\citenum{trokhymchuk:2005}, compared to simulated two-body density correlation $\rho^{(2)}_{\rm s}$ at the same volume fractions as (a--c). The color bar signifies the closest distance to a particle from the droplet interface. Note the different scales of (d), (e), and (f).}
    \label{fig:parity_plot}
\end{figure*}

We then compared $\rho^{(2)}_{\rm a}$ to $\rho^{(2)}_{\rm s}$ for different volume fractions (Fig.~\ref{fig:parity_plot}). For all volume fractions, we found acceptable agreement between the approximated and measured correlations when neither of the two particles were near the droplet interface. This agreement was expected because the particles become homogeneously distributed away from the interface, and an accurate model for $g$ for a bulk suspension should accurately predict $\rho^{(2)}$. However, significant deviations were observed for points close to the droplet interface and in suspensions with higher volume fractions. These cases exhibit more pronounced density inhomogeneities, and the approximate $\rho^{(2)}_{\rm a}$ often significantly overpredicted the two-body density correlations compared to the simulated $\rho^{(2)}_{\rm s}$.

Our analysis of the equilibrium two-body density correlations suggests a possible explanation for the failure of the DDFT calculations as drying progressed. At both $\eta_0 = 0.01$ and $\eta_0 = 0.10$, particles were initially mostly homogeneously distributed in the droplet, and we expect the two-body density correlations predicted by Eq.~\eqref{eq:twobody} to be reasonable. However, drying promotes the development of density inhomogenities due to the overall densification of the suspension and the formation of density gradients, particularly near the droplet interface. We might expect, based on Fig.~\ref{fig:parity_plot}, a poorer approximation of $\rho^{(2)}$ under these conditions. Inaccuracies in $\rho^{(2)}$ may be important for fast drying because the effective force $f$ on particles can quickly increase from its initial, equilibrium value of zero as the particles become out of equilibrium (Fig.~S10). Moreover, overestimation of the two-body density correlations might be expected to tend to enhance the contribution of $j^{(2)}$ to $j$, and we have shown that $j^{(2)}$ tends to oppose inward flux of particles. Taken together, we hence hypothesize that inaccuracies approximating $\rho^{(2)}$ cause the DDFT model to fail by ejecting particles outward from the droplet. In future, a model for $\rho^{(2)}$ that can more accurately describe inhomogeneous particle distributions may be needed to enable DDFT to describe particles suspended in droplets at larger volume fractions and farther from equilibrium.

\section{Conclusions}
\label{sec:conclusions}
In this work, we systematically explored the influence of HIs on drying-induced structure formation for hard-sphere particles suspended in a spherical droplet. Using particle-based BD simulations and continuum DDFT calculations, we analyzed the particle distributions under various drying conditions and two different approximations of the HIs: a free-draining approximation, in which particles were not hydrodynamically coupled to each other, and a pairwise far-field approximation, in which they were. For fast drying conditions (large P\'{e}clet numbers), drying produced a particle concentration gradient in all cases, as expected, due to the relatively longer time required for the particles to equilibrate by diffusion compared to the time required to dry the droplet. These concentration gradients were larger when pairwise HIs were included than for free-draining HIs; this effect originates from solvent backflow \cite{sear:pre:2017}, which resists the inward diffusion of particles and reinforces the nonuniform particle distribution. We tested for the influence of the boundary conditions on the solvent at the droplet's liquid--vapor interface---whether unbounded, slip, or no-slip---using MPCD simulations, and interestingly, found that this choice did not significantly affect the resulting particle distributions.

We developed two DDFT models for the particles in the drying droplet that leverage its radial symmetry. For free-draining HIs, the DDFT calculations were in excellent agreement with particle-based simulations, demonstrating the potential usefulness of the systematic framework provided by DDFT for constructing continuum models for drying colloidal suspensions. For pairwise far-field HIs, the DDFT calculations were in good agreement with theoretical expectations and particle-based simulations for dilute particle concentrations; however, they unexpectedly failed as the particle concentration increased, manifesting as an unphysical ejection of particles from the droplet. Particle-based simulations revealed that the likely cause of this unphysical behavior in the DDFT calculations was an inaccurate approximation of the two-body density correlations based on the bulk pair correlation function. 

Overall, our findings emphasize the critical impact of HIs on the assembly of particles confined in drying droplets. While our current methods provide key physical insights, they also point to the need for improved approaches to accurately capture the complexities of this nonequilibrium process. For example, one possible strategy to improve our DDFT model might be to compute the equilibrium two-body density correlation function self-consistently from the free-energy functional using the generalized Ornstein--Zernike equation \cite{hansen:2006}. However, this calculation would be computationally demanding because it requires evaluation of the second functional derivative of the free energy, as well as solution of an integral equation, at each timestep of the DDFT calculation. Additionally, we note that although the adiabatic assumption of DDFT seemingly works well for describing the case of free-draining HIs, superadiabatic effects may still be important to consider, particularly for the more inhomogeneous particle distributions that are produced in the presence of pairwise HIs \cite{tevrugt:2023, tevrugt:advphys:2020, delasheras:2023, schmidt:2013, schmidt:rmp:2022}.

\begin{appendix}
\section{Derivation of FMT weighted densities with radial symmetry}
\label{sec:appendix_A}
Here, we derive expressions for the weighted densities used in the FMT free-energy functional assuming radial symmetry of the particle density [Eqs.~\eqref{eq:nsphere}--\eqref{eq:ninner}]. Some of these results were previously stated in Ref.~\citenum{Roth:2010ei} for the case when $x \ge a$ [Eqs.~\eqref{eq:nsphere} and \eqref{eq:n2sphere_outer}], but here we include more details of these derivations for completeness and also extend the results to the case where $x < a$, which is required for our DDFT calculations.

To formulate expressions for $n_3$, we first explicitly apply radial symmetry in its definition,
\begin{equation}
n_3(\vv{x}) = \int \d{\vv{y}}\rho(|{\vv y}|)\theta(a-|{\vv x}-{\vv y}|).
\end{equation}
The Heaviside step function is nonzero only when the distance between $\vv{x}$ and the integration point $\vv{y}$ is less than the particle radius $a$. When $x \ge a$, this constraint sets bounds on the radial distances $y$ and the polar angles $\phi$ in spherical coordinates that contribute to $n_3$ such that
\begin{equation}
n_3(x) = \int_{x-a}^{x+a} \d{y} \, 2\pi y^2 \rho(y) \int_{\cos\phi_0}^1 \d{\cos\phi} ,
\end{equation}
where
\begin{equation}
\cos{\phi_0} = \frac{x^2+y^2-a^2}{2 x y}
\end{equation}
is the cosine of the polar angle $\phi_0$ at which the distance between two points at radial distances $x$ and $y$ is $a$. Evaluating the trivial inner integral and simplifying gives
\begin{equation}
n_3(x) = \int_{x-a}^{x+a} \d{y} \rho(y) \frac{\pi y}{x} [a^2 - (x-y)^2],
\end{equation}
which immediately leads to Eq.~\eqref{eq:nsphere} with weight function Eq.~\eqref{eq:wsphere:w3}. When $x < a$, the Heaviside step function is nonzero for all $\phi$ if $0 \le y < a - x$ but is nonzero for the same bounds on $\phi$ given above if $a - x \le y \le x + a$. The outer integral can be separated based on these two conditions to give Eq.~\eqref{eq:ninner:n3}.

To formulate expressions for $n_2$, we similarly start from
\begin{equation}
n_2(\vv{x}) = \int \d{\vv{y}} \rho(|\vv{y}|) \delta(a-|\vv{x}-\vv{y}|).
\end{equation}
In the case $x \ge a$, this integral can be evaluated using spherical coordinates and transformation of $\cos\phi$ to $u = |\vv{x}-\vv{y}|$,
\begin{align}
n_2(x) &= \int_{x-a}^{x+a} \d{y} \, 2 \pi y ^2 \rho(y) \int_{|x-y|}^{x+y} \d{u} \frac{u}{xy} \delta(a-u) \nonumber \\
&= \int_{x-a}^{x+a} \d{y} \rho(y) \frac{2 \pi y a}{x},
\end{align}
which immediately leads to Eq.~\eqref{eq:nsphere} with weight function Eq.~\eqref{eq:wsphere:w2}. When $x < a$, the argument of the Dirac delta function is always nonzero if $0 \le y < a-x$, meaning there is no contribution to $n_2$ for this condition. Restricting the limits of the outer integral accordingly to $a-x \le y \le x + a$ gives Eq.~\eqref{eq:ninner:n2}.

To formulate expressions for $\vv{n}_2$, we can use the fact that $\vv{w}_2(\vv{r}) = -\nabla_\vv{r} w_3(\vv{r})$ to compute $\vv{n}_2(\vv{x}) = -\nabla_\vv{x} n_3(\vv{x})$. Under radial symmetry, $n_3$ varies only in the radial direction so $\vv{n}_2 = -(\partial n_3/\partial x)\vv{\hat{x}}$, and the required partial derivative can be evaluated using the Leibniz integral rule. When $x \ge a$,
\begin{align}
\frac{\partial n_3}{\partial x} &= \int_{x-a}^{x+a} \d{y} \rho(y) \pi y \left[-\frac{a^2 - (x-y)^2}{x^2} -2 \frac{x-y}{x} \right] \nonumber \\
&= -\frac{n_3(x)}{x} - \int_{x-a}^{x+a} \d{y} \rho(y) \frac{2\pi y}{x} (x-y),
\label{eq:n3sphere_deriv}
\end{align}
which immediately leads to Eq.~\eqref{eq:n2sphere_outer}. Note that the second term in Eq.~\eqref{eq:n3sphere_deriv} corrects a sign error in Ref.~\citenum{Roth:2010ei}. When $x < a$,
\begin{align}
\frac{\partial n_3}{\partial x} &= -\rho(a-x) 4\pi (a-x)^2 \nonumber \\
&+ \rho(a-x) \frac{\pi (a-x)}{x} [a^2 - (2x-a)^2] \nonumber \\
&+ \int_{a-x}^{x+a} \d{y} \rho(y) \pi y \left[-\frac{a^2 - (x-y)^2}{x^2} - 2 \frac{x-y}{x} \right]
\end{align}
The first two terms sum to zero, giving Eq.~\eqref{eq:ninner:nv2}.
\end{appendix}

\section*{Supplementary Material}
See the supplementary material for particle distributions at additional initial volume fractions, additional analysis of MPCD simulations, and additional details and analysis of DDFT calculations.

\section*{Conflicts of interest}
The authors have no conflicts to disclose.

\section*{Data Availability}
The data that support the findings of this study are available from the authors upon reasonable request.

\section*{Acknowledgments}
This material is based upon work supported by the National Science Foundation under Award No.~2223084. In addition, we acknowledge funding provided by the Deutsche Forschungsgemeinschaft (DFG, German Research Foundation) through Project Nos. 405552959, 451785257, 470113688 and 509039598. This work was completed with resources provided by the Auburn University Easley Cluster and by the National High Performance Computing Center of the Dresden University of Technology.

\bibliography{references}

\begin{thebibliography}{98}%
\makeatletter
\providecommand \@ifxundefined [1]{%
 \@ifx{#1\undefined}
}%
\providecommand \@ifnum [1]{%
 \ifnum #1\expandafter \@firstoftwo
 \else \expandafter \@secondoftwo
 \fi
}%
\providecommand \@ifx [1]{%
 \ifx #1\expandafter \@firstoftwo
 \else \expandafter \@secondoftwo
 \fi
}%
\providecommand \natexlab [1]{#1}%
\providecommand \enquote  [1]{``#1''}%
\providecommand \bibnamefont  [1]{#1}%
\providecommand \bibfnamefont [1]{#1}%
\providecommand \citenamefont [1]{#1}%
\providecommand \href@noop [0]{\@secondoftwo}%
\providecommand \href [0]{\begingroup \@sanitize@url \@href}%
\providecommand \@href[1]{\@@startlink{#1}\@@href}%
\providecommand \@@href[1]{\endgroup#1\@@endlink}%
\providecommand \@sanitize@url [0]{\catcode `\\12\catcode `\$12\catcode
  `\&12\catcode `\#12\catcode `\^12\catcode `\_12\catcode `\%12\relax}%
\providecommand \@@startlink[1]{}%
\providecommand \@@endlink[0]{}%
\providecommand \url  [0]{\begingroup\@sanitize@url \@url }%
\providecommand \@url [1]{\endgroup\@href {#1}{\urlprefix }}%
\providecommand \urlprefix  [0]{URL }%
\providecommand \Eprint [0]{\href }%
\providecommand \doibase [0]{https://doi.org/}%
\providecommand \selectlanguage [0]{\@gobble}%
\providecommand \bibinfo  [0]{\@secondoftwo}%
\providecommand \bibfield  [0]{\@secondoftwo}%
\providecommand \translation [1]{[#1]}%
\providecommand \BibitemOpen [0]{}%
\providecommand \bibitemStop [0]{}%
\providecommand \bibitemNoStop [0]{.\EOS\space}%
\providecommand \EOS [0]{\spacefactor3000\relax}%
\providecommand \BibitemShut  [1]{\csname bibitem#1\endcsname}%
\let\auto@bib@innerbib\@empty
\bibitem [{\citenamefont {Walton}\ and\ \citenamefont
  {Mumford}(1999)}]{walton:1999}%
  \BibitemOpen
  \bibfield  {author} {\bibinfo {author} {\bibfnamefont {D.}~\bibnamefont
  {Walton}}\ and\ \bibinfo {author} {\bibfnamefont {C.}~\bibnamefont
  {Mumford}},\ }\bibfield  {title} {\enquote {\bibinfo {title} {Spray dried
  products—characterization of particle morphology},}\ }\href
  {https://doi.org/https://doi.org/10.1205/026387699525846} {\bibfield
  {journal} {\bibinfo  {journal} {Chem. Eng. Res. Des.}\ }\textbf {\bibinfo
  {volume} {77}},\ \bibinfo {pages} {21--38} (\bibinfo {year}
  {1999})}\BibitemShut {NoStop}%
\bibitem [{\citenamefont {Velev}, \citenamefont {Furusawa},\ and\ \citenamefont
  {Nagayama}(1996)}]{velev:1996}%
  \BibitemOpen
  \bibfield  {author} {\bibinfo {author} {\bibfnamefont {O.~D.}\ \bibnamefont
  {Velev}}, \bibinfo {author} {\bibfnamefont {K.}~\bibnamefont {Furusawa}},\
  and\ \bibinfo {author} {\bibfnamefont {K.}~\bibnamefont {Nagayama}},\
  }\bibfield  {title} {\enquote {\bibinfo {title} {Assembly of latex particles
  by using emulsion droplets as templates. 1. microstructured hollow
  spheres},}\ }\href {https://doi.org/10.1021/la9506786} {\bibfield  {journal}
  {\bibinfo  {journal} {Langmuir}\ }\textbf {\bibinfo {volume} {12}},\ \bibinfo
  {pages} {2374--2384} (\bibinfo {year} {1996})}\BibitemShut {NoStop}%
\bibitem [{\citenamefont {Rosca}, \citenamefont {Watari},\ and\ \citenamefont
  {Uo}(2004)}]{rosca:2004}%
  \BibitemOpen
  \bibfield  {author} {\bibinfo {author} {\bibfnamefont {I.~D.}\ \bibnamefont
  {Rosca}}, \bibinfo {author} {\bibfnamefont {F.}~\bibnamefont {Watari}},\ and\
  \bibinfo {author} {\bibfnamefont {M.}~\bibnamefont {Uo}},\ }\bibfield
  {title} {\enquote {\bibinfo {title} {Microparticle formation and its
  mechanism in single and double emulsion solvent evaporation},}\ }\href
  {https://doi.org/https://doi.org/10.1016/j.jconrel.2004.07.007} {\bibfield
  {journal} {\bibinfo  {journal} {J. Controlled Release}\ }\textbf {\bibinfo
  {volume} {99}},\ \bibinfo {pages} {271--280} (\bibinfo {year}
  {2004})}\BibitemShut {NoStop}%
\bibitem [{\citenamefont {Wintzheimer}\ \emph {et~al.}(2018)\citenamefont
  {Wintzheimer}, \citenamefont {Granath}, \citenamefont {Oppmann},
  \citenamefont {Kister}, \citenamefont {Thai}, \citenamefont {Kraus},
  \citenamefont {Vogel},\ and\ \citenamefont {Mandel}}]{wintzheimer:2018}%
  \BibitemOpen
  \bibfield  {author} {\bibinfo {author} {\bibfnamefont {S.}~\bibnamefont
  {Wintzheimer}}, \bibinfo {author} {\bibfnamefont {T.}~\bibnamefont
  {Granath}}, \bibinfo {author} {\bibfnamefont {M.}~\bibnamefont {Oppmann}},
  \bibinfo {author} {\bibfnamefont {T.}~\bibnamefont {Kister}}, \bibinfo
  {author} {\bibfnamefont {T.}~\bibnamefont {Thai}}, \bibinfo {author}
  {\bibfnamefont {T.}~\bibnamefont {Kraus}}, \bibinfo {author} {\bibfnamefont
  {N.}~\bibnamefont {Vogel}},\ and\ \bibinfo {author} {\bibfnamefont
  {K.}~\bibnamefont {Mandel}},\ }\bibfield  {title} {\enquote {\bibinfo {title}
  {Supraparticles: Functionality from uniform structural motifs},}\ }\href
  {https://doi.org/10.1021/acsnano.8b00873} {\bibfield  {journal} {\bibinfo
  {journal} {ACS Nano}\ }\textbf {\bibinfo {volume} {12}},\ \bibinfo {pages}
  {5093--5120} (\bibinfo {year} {2018})}\BibitemShut {NoStop}%
\bibitem [{\citenamefont {Bassani}\ \emph {et~al.}(2024)\citenamefont
  {Bassani}, \citenamefont {van Anders}, \citenamefont {Banin}, \citenamefont
  {Baranov}, \citenamefont {Chen}, \citenamefont {Dijkstra}, \citenamefont
  {Dimitriyev}, \citenamefont {Efrati}, \citenamefont {Faraudo}, \citenamefont
  {Gang}, \citenamefont {Gaston}, \citenamefont {Golestanian}, \citenamefont
  {Guerrero-Garcia}, \citenamefont {Gruenwald}, \citenamefont {Haji-Akbari},
  \citenamefont {Ibá{\~n}ez}, \citenamefont {Karg}, \citenamefont {Kraus},
  \citenamefont {Lee}, \citenamefont {Van~Lehn}, \citenamefont {Macfarlane},
  \citenamefont {Mognetti}, \citenamefont {Nikoubashman}, \citenamefont {Osat},
  \citenamefont {Prezhdo}, \citenamefont {Rotskoff}, \citenamefont {Saiz},
  \citenamefont {Shi}, \citenamefont {Skrabalak}, \citenamefont {Smalyukh},
  \citenamefont {Tagliazucchi}, \citenamefont {Talapin}, \citenamefont
  {Tkachenko}, \citenamefont {Tretiak}, \citenamefont {Vaknin}, \citenamefont
  {Widmer-Cooper}, \citenamefont {Wong}, \citenamefont {Ye}, \citenamefont
  {Zhou}, \citenamefont {Rabani}, \citenamefont {Engel},\ and\ \citenamefont
  {Travesset}}]{bassani:acsnano:2024}%
  \BibitemOpen
  \bibfield  {author} {\bibinfo {author} {\bibfnamefont {C.~L.}\ \bibnamefont
  {Bassani}}, \bibinfo {author} {\bibfnamefont {G.}~\bibnamefont {van Anders}},
  \bibinfo {author} {\bibfnamefont {U.}~\bibnamefont {Banin}}, \bibinfo
  {author} {\bibfnamefont {D.}~\bibnamefont {Baranov}}, \bibinfo {author}
  {\bibfnamefont {Q.}~\bibnamefont {Chen}}, \bibinfo {author} {\bibfnamefont
  {M.}~\bibnamefont {Dijkstra}}, \bibinfo {author} {\bibfnamefont {M.~S.}\
  \bibnamefont {Dimitriyev}}, \bibinfo {author} {\bibfnamefont
  {E.}~\bibnamefont {Efrati}}, \bibinfo {author} {\bibfnamefont
  {J.}~\bibnamefont {Faraudo}}, \bibinfo {author} {\bibfnamefont
  {O.}~\bibnamefont {Gang}}, \bibinfo {author} {\bibfnamefont {N.}~\bibnamefont
  {Gaston}}, \bibinfo {author} {\bibfnamefont {R.}~\bibnamefont {Golestanian}},
  \bibinfo {author} {\bibfnamefont {G.~I.}\ \bibnamefont {Guerrero-Garcia}},
  \bibinfo {author} {\bibfnamefont {M.}~\bibnamefont {Gruenwald}}, \bibinfo
  {author} {\bibfnamefont {A.}~\bibnamefont {Haji-Akbari}}, \bibinfo {author}
  {\bibfnamefont {M.}~\bibnamefont {Ibá{\~n}ez}}, \bibinfo {author}
  {\bibfnamefont {M.}~\bibnamefont {Karg}}, \bibinfo {author} {\bibfnamefont
  {T.}~\bibnamefont {Kraus}}, \bibinfo {author} {\bibfnamefont
  {B.}~\bibnamefont {Lee}}, \bibinfo {author} {\bibfnamefont {R.~C.}\
  \bibnamefont {Van~Lehn}}, \bibinfo {author} {\bibfnamefont {R.~J.}\
  \bibnamefont {Macfarlane}}, \bibinfo {author} {\bibfnamefont {B.~M.}\
  \bibnamefont {Mognetti}}, \bibinfo {author} {\bibfnamefont {A.}~\bibnamefont
  {Nikoubashman}}, \bibinfo {author} {\bibfnamefont {S.}~\bibnamefont {Osat}},
  \bibinfo {author} {\bibfnamefont {O.~V.}\ \bibnamefont {Prezhdo}}, \bibinfo
  {author} {\bibfnamefont {G.~M.}\ \bibnamefont {Rotskoff}}, \bibinfo {author}
  {\bibfnamefont {L.}~\bibnamefont {Saiz}}, \bibinfo {author} {\bibfnamefont
  {A.-C.}\ \bibnamefont {Shi}}, \bibinfo {author} {\bibfnamefont
  {S.}~\bibnamefont {Skrabalak}}, \bibinfo {author} {\bibfnamefont {I.~I.}\
  \bibnamefont {Smalyukh}}, \bibinfo {author} {\bibfnamefont {M.}~\bibnamefont
  {Tagliazucchi}}, \bibinfo {author} {\bibfnamefont {D.~V.}\ \bibnamefont
  {Talapin}}, \bibinfo {author} {\bibfnamefont {A.~V.}\ \bibnamefont
  {Tkachenko}}, \bibinfo {author} {\bibfnamefont {S.}~\bibnamefont {Tretiak}},
  \bibinfo {author} {\bibfnamefont {D.}~\bibnamefont {Vaknin}}, \bibinfo
  {author} {\bibfnamefont {A.}~\bibnamefont {Widmer-Cooper}}, \bibinfo {author}
  {\bibfnamefont {G.~C.~L.}\ \bibnamefont {Wong}}, \bibinfo {author}
  {\bibfnamefont {X.}~\bibnamefont {Ye}}, \bibinfo {author} {\bibfnamefont
  {S.}~\bibnamefont {Zhou}}, \bibinfo {author} {\bibfnamefont {E.}~\bibnamefont
  {Rabani}}, \bibinfo {author} {\bibfnamefont {M.}~\bibnamefont {Engel}},\ and\
  \bibinfo {author} {\bibfnamefont {A.}~\bibnamefont {Travesset}},\ }\bibfield
  {title} {\enquote {\bibinfo {title} {Nanocrystal assemblies: Current advances
  and open problems},}\ }\href@noop {} {\bibfield  {journal} {\bibinfo
  {journal} {ACS Nano}\ }\textbf {\bibinfo {volume} {18}},\ \bibinfo {pages}
  {14791--14840} (\bibinfo {year} {2024})}\BibitemShut {NoStop}%
\bibitem [{\citenamefont {Iskandar}, \citenamefont {Chang},\ and\ \citenamefont
  {Okuyama}(2003)}]{iskandar:2003}%
  \BibitemOpen
  \bibfield  {author} {\bibinfo {author} {\bibfnamefont {F.}~\bibnamefont
  {Iskandar}}, \bibinfo {author} {\bibfnamefont {H.}~\bibnamefont {Chang}},\
  and\ \bibinfo {author} {\bibfnamefont {K.}~\bibnamefont {Okuyama}},\
  }\bibfield  {title} {\enquote {\bibinfo {title} {Preparation of
  microencapsulated powders by an aerosol spray method and their optical
  properties},}\ }\href
  {https://doi.org/https://doi.org/10.1163/15685520360685983} {\bibfield
  {journal} {\bibinfo  {journal} {Adv. Powder Technol.}\ }\textbf {\bibinfo
  {volume} {14}},\ \bibinfo {pages} {349--367} (\bibinfo {year}
  {2003})}\BibitemShut {NoStop}%
\bibitem [{\citenamefont {Wang}\ \emph {et~al.}(2020)\citenamefont {Wang},
  \citenamefont {Sultan}, \citenamefont {Goerlitzer}, \citenamefont {Mbah},
  \citenamefont {Engel},\ and\ \citenamefont {Vogel}}]{wang:2020}%
  \BibitemOpen
  \bibfield  {author} {\bibinfo {author} {\bibfnamefont {J.}~\bibnamefont
  {Wang}}, \bibinfo {author} {\bibfnamefont {U.}~\bibnamefont {Sultan}},
  \bibinfo {author} {\bibfnamefont {E.~S.~A.}\ \bibnamefont {Goerlitzer}},
  \bibinfo {author} {\bibfnamefont {C.~F.}\ \bibnamefont {Mbah}}, \bibinfo
  {author} {\bibfnamefont {M.}~\bibnamefont {Engel}},\ and\ \bibinfo {author}
  {\bibfnamefont {N.}~\bibnamefont {Vogel}},\ }\bibfield  {title} {\enquote
  {\bibinfo {title} {Structural color of colloidal clusters as a tool to
  investigate structure and dynamics},}\ }\href
  {https://doi.org/https://doi.org/10.1002/adfm.201907730} {\bibfield
  {journal} {\bibinfo  {journal} {Adv. Funct. Mater.}\ }\textbf {\bibinfo
  {volume} {30}},\ \bibinfo {pages} {1907730} (\bibinfo {year}
  {2020})}\BibitemShut {NoStop}%
\bibitem [{\citenamefont {Luo}\ \emph {et~al.}(2014)\citenamefont {Luo},
  \citenamefont {Zhang}, \citenamefont {Sun}, \citenamefont {Chu},
  \citenamefont {Zhou}, \citenamefont {Guo}, \citenamefont {Chen},\ and\
  \citenamefont {Xu}}]{luo:2014}%
  \BibitemOpen
  \bibfield  {author} {\bibinfo {author} {\bibfnamefont {Y.}~\bibnamefont
  {Luo}}, \bibinfo {author} {\bibfnamefont {J.}~\bibnamefont {Zhang}}, \bibinfo
  {author} {\bibfnamefont {A.}~\bibnamefont {Sun}}, \bibinfo {author}
  {\bibfnamefont {C.}~\bibnamefont {Chu}}, \bibinfo {author} {\bibfnamefont
  {S.}~\bibnamefont {Zhou}}, \bibinfo {author} {\bibfnamefont {J.}~\bibnamefont
  {Guo}}, \bibinfo {author} {\bibfnamefont {T.}~\bibnamefont {Chen}},\ and\
  \bibinfo {author} {\bibfnamefont {G.}~\bibnamefont {Xu}},\ }\bibfield
  {title} {\enquote {\bibinfo {title} {Electric field induced structural color
  changes of sio2@tio2 core–shell colloidal suspensions},}\ }\href
  {https://doi.org/10.1039/C3TC32227K} {\bibfield  {journal} {\bibinfo
  {journal} {J. Mater. Chem. C}\ }\textbf {\bibinfo {volume} {2}},\ \bibinfo
  {pages} {1990--1994} (\bibinfo {year} {2014})}\BibitemShut {NoStop}%
\bibitem [{\citenamefont {Patil}\ \emph {et~al.}(2022)\citenamefont {Patil},
  \citenamefont {Heil}, \citenamefont {Vanthournout}, \citenamefont {Bleuel},
  \citenamefont {Singla}, \citenamefont {Hu}, \citenamefont {Gianneschi},
  \citenamefont {Shawkey}, \citenamefont {Sinha}, \citenamefont {Jayaraman},\
  and\ \citenamefont {Dhinojwala}}]{patil:aom:2022}%
  \BibitemOpen
  \bibfield  {author} {\bibinfo {author} {\bibfnamefont {A.}~\bibnamefont
  {Patil}}, \bibinfo {author} {\bibfnamefont {C.~M.}\ \bibnamefont {Heil}},
  \bibinfo {author} {\bibfnamefont {B.}~\bibnamefont {Vanthournout}}, \bibinfo
  {author} {\bibfnamefont {M.}~\bibnamefont {Bleuel}}, \bibinfo {author}
  {\bibfnamefont {S.}~\bibnamefont {Singla}}, \bibinfo {author} {\bibfnamefont
  {Z.}~\bibnamefont {Hu}}, \bibinfo {author} {\bibfnamefont {N.~C.}\
  \bibnamefont {Gianneschi}}, \bibinfo {author} {\bibfnamefont {M.~D.}\
  \bibnamefont {Shawkey}}, \bibinfo {author} {\bibfnamefont {S.~K.}\
  \bibnamefont {Sinha}}, \bibinfo {author} {\bibfnamefont {A.}~\bibnamefont
  {Jayaraman}},\ and\ \bibinfo {author} {\bibfnamefont {A.}~\bibnamefont
  {Dhinojwala}},\ }\bibfield  {title} {\enquote {\bibinfo {title} {Structural
  color production in melanin-based disordered colloidal nanoparticle
  assemblies in spherical confinement},}\ }\href@noop {} {\bibfield  {journal}
  {\bibinfo  {journal} {Adv. Opt. Mater.}\ }\textbf {\bibinfo {volume} {10}},\
  \bibinfo {pages} {2102162} (\bibinfo {year} {2022})}\BibitemShut {NoStop}%
\bibitem [{\citenamefont {Heil}\ \emph {et~al.}(2023)\citenamefont {Heil},
  \citenamefont {Patil}, \citenamefont {Vanthournout}, \citenamefont {Bleuel},
  \citenamefont {Song}, \citenamefont {Hu}, \citenamefont {Gianneschi},
  \citenamefont {Shawkey}, \citenamefont {Sinha}, \citenamefont {Jayaraman},\
  and\ \citenamefont {Dhinojwala}}]{heil:scadv:2023}%
  \BibitemOpen
  \bibfield  {author} {\bibinfo {author} {\bibfnamefont {C.~M.}\ \bibnamefont
  {Heil}}, \bibinfo {author} {\bibfnamefont {A.}~\bibnamefont {Patil}},
  \bibinfo {author} {\bibfnamefont {B.}~\bibnamefont {Vanthournout}}, \bibinfo
  {author} {\bibfnamefont {M.}~\bibnamefont {Bleuel}}, \bibinfo {author}
  {\bibfnamefont {J.-J.}\ \bibnamefont {Song}}, \bibinfo {author}
  {\bibfnamefont {Z.}~\bibnamefont {Hu}}, \bibinfo {author} {\bibfnamefont
  {N.~C.}\ \bibnamefont {Gianneschi}}, \bibinfo {author} {\bibfnamefont
  {M.~D.}\ \bibnamefont {Shawkey}}, \bibinfo {author} {\bibfnamefont {S.~K.}\
  \bibnamefont {Sinha}}, \bibinfo {author} {\bibfnamefont {A.}~\bibnamefont
  {Jayaraman}},\ and\ \bibinfo {author} {\bibfnamefont {A.}~\bibnamefont
  {Dhinojwala}},\ }\bibfield  {title} {\enquote {\bibinfo {title} {Mechanism of
  structural colors in binary mixtures of nanoparticle-based supraballs},}\
  }\href@noop {} {\bibfield  {journal} {\bibinfo  {journal} {Sci. Adv.}\
  }\textbf {\bibinfo {volume} {9}},\ \bibinfo {pages} {eadf2859} (\bibinfo
  {year} {2023})}\BibitemShut {NoStop}%
\bibitem [{\citenamefont {Gradon}\ \emph {et~al.}(2020)\citenamefont {Gradon},
  \citenamefont {Balgis}, \citenamefont {Hirano}, \citenamefont {Rahmatika},
  \citenamefont {Ogi},\ and\ \citenamefont {Okuyama}}]{leon:2020}%
  \BibitemOpen
  \bibfield  {author} {\bibinfo {author} {\bibfnamefont {L.}~\bibnamefont
  {Gradon}}, \bibinfo {author} {\bibfnamefont {R.}~\bibnamefont {Balgis}},
  \bibinfo {author} {\bibfnamefont {T.}~\bibnamefont {Hirano}}, \bibinfo
  {author} {\bibfnamefont {A.~M.}\ \bibnamefont {Rahmatika}}, \bibinfo {author}
  {\bibfnamefont {T.}~\bibnamefont {Ogi}},\ and\ \bibinfo {author}
  {\bibfnamefont {K.}~\bibnamefont {Okuyama}},\ }\bibfield  {title} {\enquote
  {\bibinfo {title} {Advanced aerosol technologies towards structure and
  morphologically controlled next-generation catalytic materials},}\ }\href
  {https://doi.org/https://doi.org/10.1016/j.jaerosci.2020.105608} {\bibfield
  {journal} {\bibinfo  {journal} {J. Aerosol Sci.}\ }\textbf {\bibinfo {volume}
  {149}},\ \bibinfo {pages} {105608} (\bibinfo {year} {2020})}\BibitemShut
  {NoStop}%
\bibitem [{\citenamefont {Hou}, \citenamefont {Han},\ and\ \citenamefont
  {Tang}(2020)}]{hou:2020}%
  \BibitemOpen
  \bibfield  {author} {\bibinfo {author} {\bibfnamefont {K.}~\bibnamefont
  {Hou}}, \bibinfo {author} {\bibfnamefont {J.}~\bibnamefont {Han}},\ and\
  \bibinfo {author} {\bibfnamefont {Z.}~\bibnamefont {Tang}},\ }\bibfield
  {title} {\enquote {\bibinfo {title} {Formation of supraparticles and their
  application in catalysis},}\ }\href
  {https://doi.org/10.1021/acsmaterialslett.9b00446} {\bibfield  {journal}
  {\bibinfo  {journal} {ACS Mater. Lett.}\ }\textbf {\bibinfo {volume} {2}},\
  \bibinfo {pages} {95--106} (\bibinfo {year} {2020})}\BibitemShut {NoStop}%
\bibitem [{\citenamefont {Bodmeier}\ and\ \citenamefont
  {McGinity}(1987)}]{bodmeier:1987}%
  \BibitemOpen
  \bibfield  {author} {\bibinfo {author} {\bibfnamefont {R.}~\bibnamefont
  {Bodmeier}}\ and\ \bibinfo {author} {\bibfnamefont {J.~W.}\ \bibnamefont
  {McGinity}},\ }\bibfield  {title} {\enquote {\bibinfo {title} {The
  preparation and evaluation of drug-containing poly (dl-lactide) microspheres
  formed by the solvent evaporation method},}\ }\href@noop {} {\bibfield
  {journal} {\bibinfo  {journal} {Pharm. Res.}\ }\textbf {\bibinfo {volume}
  {4}},\ \bibinfo {pages} {465--471} (\bibinfo {year} {1987})}\BibitemShut
  {NoStop}%
\bibitem [{\citenamefont {De~Lima}\ \emph {et~al.}(2023)\citenamefont
  {De~Lima}, \citenamefont {Matos}, \citenamefont {De~S{\'a}}, \citenamefont
  {Mashiba}, \citenamefont {Magalh{\~a}es}, \citenamefont {Rachwal},\ and\
  \citenamefont {Zanatta}}]{de_lima:2023}%
  \BibitemOpen
  \bibfield  {author} {\bibinfo {author} {\bibfnamefont {G.~G.}\ \bibnamefont
  {De~Lima}}, \bibinfo {author} {\bibfnamefont {M.}~\bibnamefont {Matos}},
  \bibinfo {author} {\bibfnamefont {F.~P.}\ \bibnamefont {De~S{\'a}}}, \bibinfo
  {author} {\bibfnamefont {L.~N.}\ \bibnamefont {Mashiba}}, \bibinfo {author}
  {\bibfnamefont {W.~L.}\ \bibnamefont {Magalh{\~a}es}}, \bibinfo {author}
  {\bibfnamefont {M.~F.~G.}\ \bibnamefont {Rachwal}},\ and\ \bibinfo {author}
  {\bibfnamefont {J.~A.}\ \bibnamefont {Zanatta}},\ }\bibfield  {title}
  {\enquote {\bibinfo {title} {Supraparticles as slow-release fertiliser in
  seedling potential growth of eucalyptus urograndis and greenhouse gas flux
  impacts},}\ }\href
  {https://doi.org/https://doi.org/10.1007/s11356-022-23845-5} {\bibfield
  {journal} {\bibinfo  {journal} {Environ. Sci. Pollut. Res.}\ }\textbf
  {\bibinfo {volume} {30}},\ \bibinfo {pages} {23047--23059} (\bibinfo {year}
  {2023})}\BibitemShut {NoStop}%
\bibitem [{\citenamefont {Liu}\ \emph {et~al.}(2019)\citenamefont {Liu},
  \citenamefont {Midya}, \citenamefont {Kappl}, \citenamefont {Butt},\ and\
  \citenamefont {Nikoubashman}}]{liu:march:2019}%
  \BibitemOpen
  \bibfield  {author} {\bibinfo {author} {\bibfnamefont {W.}~\bibnamefont
  {Liu}}, \bibinfo {author} {\bibfnamefont {J.}~\bibnamefont {Midya}}, \bibinfo
  {author} {\bibfnamefont {M.}~\bibnamefont {Kappl}}, \bibinfo {author}
  {\bibfnamefont {H.-J.}\ \bibnamefont {Butt}},\ and\ \bibinfo {author}
  {\bibfnamefont {A.}~\bibnamefont {Nikoubashman}},\ }\bibfield  {title}
  {\enquote {\bibinfo {title} {Segregation in drying binary colloidal
  droplets},}\ }\href {https://doi.org/10.1021/acsnano.9b00459} {\bibfield
  {journal} {\bibinfo  {journal} {ACS Nano}\ }\textbf {\bibinfo {volume}
  {13}},\ \bibinfo {pages} {4972--4979} (\bibinfo {year} {2019})}\BibitemShut
  {NoStop}%
\bibitem [{\citenamefont {Liu}, \citenamefont {Kappl},\ and\ \citenamefont
  {Butt}(2019)}]{liu:december:2019}%
  \BibitemOpen
  \bibfield  {author} {\bibinfo {author} {\bibfnamefont {W.}~\bibnamefont
  {Liu}}, \bibinfo {author} {\bibfnamefont {M.}~\bibnamefont {Kappl}},\ and\
  \bibinfo {author} {\bibfnamefont {H.-J.}\ \bibnamefont {Butt}},\ }\bibfield
  {title} {\enquote {\bibinfo {title} {Tuning the porosity of
  supraparticles},}\ }\href {https://doi.org/10.1021/acsnano.9b05673}
  {\bibfield  {journal} {\bibinfo  {journal} {ACS Nano}\ }\textbf {\bibinfo
  {volume} {13}},\ \bibinfo {pages} {13949--13956} (\bibinfo {year}
  {2019})}\BibitemShut {NoStop}%
\bibitem [{\citenamefont {Liu}\ \emph {et~al.}(2022)\citenamefont {Liu},
  \citenamefont {Kappl}, \citenamefont {Steffen},\ and\ \citenamefont
  {Butt}}]{liu:2022}%
  \BibitemOpen
  \bibfield  {author} {\bibinfo {author} {\bibfnamefont {W.}~\bibnamefont
  {Liu}}, \bibinfo {author} {\bibfnamefont {M.}~\bibnamefont {Kappl}}, \bibinfo
  {author} {\bibfnamefont {W.}~\bibnamefont {Steffen}},\ and\ \bibinfo {author}
  {\bibfnamefont {H.-J.}\ \bibnamefont {Butt}},\ }\bibfield  {title} {\enquote
  {\bibinfo {title} {Controlling supraparticle shape and structure by tuning
  colloidal interactions},}\ }\href
  {https://doi.org/https://doi.org/10.1016/j.jcis.2021.09.035} {\bibfield
  {journal} {\bibinfo  {journal} {J. Colloid Interface Sci.}\ }\textbf
  {\bibinfo {volume} {607}},\ \bibinfo {pages} {1661--1670} (\bibinfo {year}
  {2022})}\BibitemShut {NoStop}%
\bibitem [{\citenamefont {Routh}\ and\ \citenamefont
  {B~Zimmerman}(2004)}]{routh:2004}%
  \BibitemOpen
  \bibfield  {author} {\bibinfo {author} {\bibfnamefont {A.~F.}\ \bibnamefont
  {Routh}}\ and\ \bibinfo {author} {\bibfnamefont {W.}~\bibnamefont
  {B~Zimmerman}},\ }\bibfield  {title} {\enquote {\bibinfo {title}
  {Distribution of particles during solvent evaporation from films},}\ }\href
  {https://doi.org/https://doi.org/10.1016/j.ces.2004.04.027} {\bibfield
  {journal} {\bibinfo  {journal} {Chem. Eng. Sci.}\ }\textbf {\bibinfo {volume}
  {59}},\ \bibinfo {pages} {2961--2968} (\bibinfo {year} {2004})}\BibitemShut
  {NoStop}%
\bibitem [{\citenamefont {Howard}, \citenamefont {Nikoubashman},\ and\
  \citenamefont {Panagiotopoulos}(2017{\natexlab{a}})}]{howard:lng:2017}%
  \BibitemOpen
  \bibfield  {author} {\bibinfo {author} {\bibfnamefont {M.~P.}\ \bibnamefont
  {Howard}}, \bibinfo {author} {\bibfnamefont {A.}~\bibnamefont
  {Nikoubashman}},\ and\ \bibinfo {author} {\bibfnamefont {A.~Z.}\ \bibnamefont
  {Panagiotopoulos}},\ }\bibfield  {title} {\enquote {\bibinfo {title}
  {Stratification dynamics in drying colloidal mixtures},}\ }\href@noop {}
  {\bibfield  {journal} {\bibinfo  {journal} {Langmuir}\ }\textbf {\bibinfo
  {volume} {33}},\ \bibinfo {pages} {3685--3693} (\bibinfo {year}
  {2017}{\natexlab{a}})}\BibitemShut {NoStop}%
\bibitem [{\citenamefont {Tang}, \citenamefont {Grest},\ and\ \citenamefont
  {Cheng}(2019)}]{tang:jcp:2019}%
  \BibitemOpen
  \bibfield  {author} {\bibinfo {author} {\bibfnamefont {Y.}~\bibnamefont
  {Tang}}, \bibinfo {author} {\bibfnamefont {G.~S.}\ \bibnamefont {Grest}},\
  and\ \bibinfo {author} {\bibfnamefont {S.}~\bibnamefont {Cheng}},\ }\bibfield
   {title} {\enquote {\bibinfo {title} {Stratification of drying particle
  suspensions: Comparison of implicit and explicit solvent simulations},}\
  }\href {https://doi.org/10.1063/1.5066035} {\bibfield  {journal} {\bibinfo
  {journal} {J. Chem. Phys.}\ }\textbf {\bibinfo {volume} {150}},\ \bibinfo
  {pages} {224901} (\bibinfo {year} {2019})}\BibitemShut {NoStop}%
\bibitem [{\citenamefont {Yamaguchi}, \citenamefont {Kimura},\ and\
  \citenamefont {Hirota}(1998)}]{yamaguchi:1998}%
  \BibitemOpen
  \bibfield  {author} {\bibinfo {author} {\bibfnamefont {T.}~\bibnamefont
  {Yamaguchi}}, \bibinfo {author} {\bibfnamefont {Y.}~\bibnamefont {Kimura}},\
  and\ \bibinfo {author} {\bibfnamefont {N.}~\bibnamefont {Hirota}},\
  }\bibfield  {title} {\enquote {\bibinfo {title} {Molecular dynamics
  simulation of solute diffusion in lennard-jones fluids},}\ }\href
  {https://doi.org/10.1080/002689798168033} {\bibfield  {journal} {\bibinfo
  {journal} {Mol. Phys.}\ }\textbf {\bibinfo {volume} {94}},\ \bibinfo {pages}
  {527--537} (\bibinfo {year} {1998})}\BibitemShut {NoStop}%
\bibitem [{\citenamefont {Chen}, \citenamefont {Koplik},\ and\ \citenamefont
  {Kretzschmar}(2013)}]{chen:2013}%
  \BibitemOpen
  \bibfield  {author} {\bibinfo {author} {\bibfnamefont {W.}~\bibnamefont
  {Chen}}, \bibinfo {author} {\bibfnamefont {J.}~\bibnamefont {Koplik}},\ and\
  \bibinfo {author} {\bibfnamefont {I.}~\bibnamefont {Kretzschmar}},\
  }\bibfield  {title} {\enquote {\bibinfo {title} {Molecular dynamics
  simulations of the evaporation of particle-laden droplets},}\ }\href
  {https://doi.org/10.1103/PhysRevE.87.052404} {\bibfield  {journal} {\bibinfo
  {journal} {Phys. Rev. E}\ }\textbf {\bibinfo {volume} {87}},\ \bibinfo
  {pages} {052404} (\bibinfo {year} {2013})}\BibitemShut {NoStop}%
\bibitem [{\citenamefont {Chen}\ and\ \citenamefont
  {Doolen}(1998)}]{chen:1998}%
  \BibitemOpen
  \bibfield  {author} {\bibinfo {author} {\bibfnamefont {S.}~\bibnamefont
  {Chen}}\ and\ \bibinfo {author} {\bibfnamefont {G.~D.}\ \bibnamefont
  {Doolen}},\ }\bibfield  {title} {\enquote {\bibinfo {title} {Lattice
  boltzmann method for fluid flows},}\ }\href
  {https://doi.org/https://doi.org/10.1146/annurev.fluid.30.1.329} {\bibfield
  {journal} {\bibinfo  {journal} {Annu. Rev. Fluid Mech.}\ }\textbf {\bibinfo
  {volume} {30}},\ \bibinfo {pages} {329--364} (\bibinfo {year}
  {1998})}\BibitemShut {NoStop}%
\bibitem [{\citenamefont {Malevanets}\ and\ \citenamefont
  {Kapral}(1999)}]{malevanets:1999}%
  \BibitemOpen
  \bibfield  {author} {\bibinfo {author} {\bibfnamefont {A.}~\bibnamefont
  {Malevanets}}\ and\ \bibinfo {author} {\bibfnamefont {R.}~\bibnamefont
  {Kapral}},\ }\bibfield  {title} {\enquote {\bibinfo {title} {{Mesoscopic
  model for solvent dynamics}},}\ }\href {https://doi.org/10.1063/1.478857}
  {\bibfield  {journal} {\bibinfo  {journal} {J. Chem. Phys.}\ }\textbf
  {\bibinfo {volume} {110}},\ \bibinfo {pages} {8605--8613} (\bibinfo {year}
  {1999})}\BibitemShut {NoStop}%
\bibitem [{\citenamefont {Brady}\ \emph {et~al.}(1988)\citenamefont {Brady},
  \citenamefont {Phillips}, \citenamefont {Lester},\ and\ \citenamefont
  {Bossis}}]{brady:jfm:1988}%
  \BibitemOpen
  \bibfield  {author} {\bibinfo {author} {\bibfnamefont {J.~F.}\ \bibnamefont
  {Brady}}, \bibinfo {author} {\bibfnamefont {R.~J.}\ \bibnamefont {Phillips}},
  \bibinfo {author} {\bibfnamefont {J.~C.}\ \bibnamefont {Lester}},\ and\
  \bibinfo {author} {\bibfnamefont {G.}~\bibnamefont {Bossis}},\ }\bibfield
  {title} {\enquote {\bibinfo {title} {Dynamic simulation of hydrodynamically
  interacting suspensions},}\ }\href
  {https://doi.org/10.1017/S0022112088002411} {\bibfield  {journal} {\bibinfo
  {journal} {J. Fluid Mech.}\ }\textbf {\bibinfo {volume} {195}},\ \bibinfo
  {pages} {257–280} (\bibinfo {year} {1988})}\BibitemShut {NoStop}%
\bibitem [{\citenamefont {Routh}(2013)}]{routh:2013}%
  \BibitemOpen
  \bibfield  {author} {\bibinfo {author} {\bibfnamefont {A.~F.}\ \bibnamefont
  {Routh}},\ }\bibfield  {title} {\enquote {\bibinfo {title} {Drying of thin
  colloidal films},}\ }\href {https://doi.org/10.1088/0034-4885/76/4/046603}
  {\bibfield  {journal} {\bibinfo  {journal} {Rep. Prog. Phys.}\ }\textbf
  {\bibinfo {volume} {76}},\ \bibinfo {pages} {046603} (\bibinfo {year}
  {2013})}\BibitemShut {NoStop}%
\bibitem [{\citenamefont {Sear}\ and\ \citenamefont
  {Warren}(2017)}]{sear:pre:2017}%
  \BibitemOpen
  \bibfield  {author} {\bibinfo {author} {\bibfnamefont {R.~P.}\ \bibnamefont
  {Sear}}\ and\ \bibinfo {author} {\bibfnamefont {P.~B.}\ \bibnamefont
  {Warren}},\ }\bibfield  {title} {\enquote {\bibinfo {title} {Diffusiophoresis
  in nonadsorbing polymer solutions: The asakura-oosawa model and
  stratification in drying films},}\ }\href
  {https://doi.org/10.1103/PhysRevE.96.062602} {\bibfield  {journal} {\bibinfo
  {journal} {Phys. Rev. E}\ }\textbf {\bibinfo {volume} {96}},\ \bibinfo
  {pages} {062602} (\bibinfo {year} {2017})}\BibitemShut {NoStop}%
\bibitem [{\citenamefont {Zhou}, \citenamefont {Jiang},\ and\ \citenamefont
  {Doi}(2017)}]{zhou:prl:2017}%
  \BibitemOpen
  \bibfield  {author} {\bibinfo {author} {\bibfnamefont {J.}~\bibnamefont
  {Zhou}}, \bibinfo {author} {\bibfnamefont {Y.}~\bibnamefont {Jiang}},\ and\
  \bibinfo {author} {\bibfnamefont {M.}~\bibnamefont {Doi}},\ }\bibfield
  {title} {\enquote {\bibinfo {title} {Cross interaction drives stratification
  in drying film of binary colloidal mixtures},}\ }\href
  {https://doi.org/10.1103/PhysRevLett.118.108002} {\bibfield  {journal}
  {\bibinfo  {journal} {Phys. Rev. Lett.}\ }\textbf {\bibinfo {volume} {118}},\
  \bibinfo {pages} {108002} (\bibinfo {year} {2017})}\BibitemShut {NoStop}%
\bibitem [{\citenamefont {Chun}, \citenamefont {Yoo},\ and\ \citenamefont
  {Jung}(2020)}]{chun:2020}%
  \BibitemOpen
  \bibfield  {author} {\bibinfo {author} {\bibfnamefont {B.}~\bibnamefont
  {Chun}}, \bibinfo {author} {\bibfnamefont {T.}~\bibnamefont {Yoo}},\ and\
  \bibinfo {author} {\bibfnamefont {H.~W.}\ \bibnamefont {Jung}},\ }\bibfield
  {title} {\enquote {\bibinfo {title} {Temporal evolution of concentration and
  microstructure of colloidal films during vertical drying: a lattice boltzmann
  simulation study},}\ }\href {https://doi.org/10.1039/C9SM01925A} {\bibfield
  {journal} {\bibinfo  {journal} {Soft Matter}\ }\textbf {\bibinfo {volume}
  {16}},\ \bibinfo {pages} {523--533} (\bibinfo {year} {2020})}\BibitemShut
  {NoStop}%
\bibitem [{\citenamefont {He}\ \emph {et~al.}(2021)\citenamefont {He},
  \citenamefont {Martin-Fabiani}, \citenamefont {Roth}, \citenamefont
  {T\'{o}th},\ and\ \citenamefont {Archer}}]{he:lng:2021}%
  \BibitemOpen
  \bibfield  {author} {\bibinfo {author} {\bibfnamefont {B.}~\bibnamefont
  {He}}, \bibinfo {author} {\bibfnamefont {I.}~\bibnamefont {Martin-Fabiani}},
  \bibinfo {author} {\bibfnamefont {R.}~\bibnamefont {Roth}}, \bibinfo {author}
  {\bibfnamefont {G.~I.}\ \bibnamefont {T\'{o}th}},\ and\ \bibinfo {author}
  {\bibfnamefont {A.~J.}\ \bibnamefont {Archer}},\ }\bibfield  {title}
  {\enquote {\bibinfo {title} {Dynamical density functional theory for the
  drying and stratification of binary colloidal dispersions},}\ }\href
  {https://doi.org/10.1021/acs.langmuir.0c02825} {\bibfield  {journal}
  {\bibinfo  {journal} {Langmuir}\ }\textbf {\bibinfo {volume} {37}},\ \bibinfo
  {pages} {1399--1409} (\bibinfo {year} {2021})}\BibitemShut {NoStop}%
\bibitem [{\citenamefont {Rees-Zimmerman}\ and\ \citenamefont
  {Routh}(2021)}]{rees-zimmerman:2021}%
  \BibitemOpen
  \bibfield  {author} {\bibinfo {author} {\bibfnamefont {C.~R.}\ \bibnamefont
  {Rees-Zimmerman}}\ and\ \bibinfo {author} {\bibfnamefont {A.~F.}\
  \bibnamefont {Routh}},\ }\bibfield  {title} {\enquote {\bibinfo {title}
  {Stratification in drying films: a diffusion–diffusiophoresis model},}\
  }\href {https://doi.org/10.1017/jfm.2021.800} {\bibfield  {journal} {\bibinfo
   {journal} {J. Fluid Mech.}\ }\textbf {\bibinfo {volume} {928}},\ \bibinfo
  {pages} {A15} (\bibinfo {year} {2021})}\BibitemShut {NoStop}%
\bibitem [{\citenamefont {Yoo}, \citenamefont {Chun},\ and\ \citenamefont
  {Jung}(2022)}]{yoo:2022}%
  \BibitemOpen
  \bibfield  {author} {\bibinfo {author} {\bibfnamefont {T.}~\bibnamefont
  {Yoo}}, \bibinfo {author} {\bibfnamefont {B.}~\bibnamefont {Chun}},\ and\
  \bibinfo {author} {\bibfnamefont {H.~W.}\ \bibnamefont {Jung}},\ }\bibfield
  {title} {\enquote {\bibinfo {title} {Practical drying model for horizontal
  colloidal films in rapid evaporation processes},}\ }\href
  {https://doi.org/10.1080/07373937.2020.1811723} {\bibfield  {journal}
  {\bibinfo  {journal} {Drying Technol.}\ }\textbf {\bibinfo {volume} {40}},\
  \bibinfo {pages} {516--526} (\bibinfo {year} {2022})}\BibitemShut {NoStop}%
\bibitem [{\citenamefont {Kundu}\ and\ \citenamefont
  {Howard}(2022)}]{kundu:2022}%
  \BibitemOpen
  \bibfield  {author} {\bibinfo {author} {\bibfnamefont {M.}~\bibnamefont
  {Kundu}}\ and\ \bibinfo {author} {\bibfnamefont {M.~P.}\ \bibnamefont
  {Howard}},\ }\bibfield  {title} {\enquote {\bibinfo {title} {{Dynamic density
  functional theory for drying colloidal suspensions: Comparison of hard-sphere
  free-energy functionals}},}\ }\href {https://doi.org/10.1063/5.0118695}
  {\bibfield  {journal} {\bibinfo  {journal} {J. Chem. Phys.}\ }\textbf
  {\bibinfo {volume} {157}},\ \bibinfo {pages} {184904} (\bibinfo {year}
  {2022})}\BibitemShut {NoStop}%
\bibitem [{\citenamefont {Brady}(2011)}]{brady:2011bh}%
  \BibitemOpen
  \bibfield  {author} {\bibinfo {author} {\bibfnamefont {J.~F.}\ \bibnamefont
  {Brady}},\ }\bibfield  {title} {\enquote {\bibinfo {title} {Particle motion
  driven by solute gradients with application to autonomous motion: continuum
  and colloidal perspectives},}\ }\href@noop {} {\bibfield  {journal} {\bibinfo
   {journal} {J. Fluid Mech.}\ }\textbf {\bibinfo {volume} {667}},\ \bibinfo
  {pages} {216--259} (\bibinfo {year} {2011})}\BibitemShut {NoStop}%
\bibitem [{\citenamefont {Statt}, \citenamefont {Howard},\ and\ \citenamefont
  {Panagiotopoulos}(2018)}]{antonia:2018}%
  \BibitemOpen
  \bibfield  {author} {\bibinfo {author} {\bibfnamefont {A.}~\bibnamefont
  {Statt}}, \bibinfo {author} {\bibfnamefont {M.~P.}\ \bibnamefont {Howard}},\
  and\ \bibinfo {author} {\bibfnamefont {A.~Z.}\ \bibnamefont
  {Panagiotopoulos}},\ }\bibfield  {title} {\enquote {\bibinfo {title}
  {{Influence of hydrodynamic interactions on stratification in drying
  mixtures}},}\ }\href {https://doi.org/10.1063/1.5031789} {\bibfield
  {journal} {\bibinfo  {journal} {J. Chem. Phys.}\ }\textbf {\bibinfo {volume}
  {149}},\ \bibinfo {pages} {024902} (\bibinfo {year} {2018})}\BibitemShut
  {NoStop}%
\bibitem [{\citenamefont {Howard}\ \emph {et~al.}(2018)\citenamefont {Howard},
  \citenamefont {Reinhart}, \citenamefont {Sanyal}, \citenamefont {Shell},
  \citenamefont {Nikoubashman},\ and\ \citenamefont
  {Panagiotopoulos}}]{howard:jcp:2018}%
  \BibitemOpen
  \bibfield  {author} {\bibinfo {author} {\bibfnamefont {M.~P.}\ \bibnamefont
  {Howard}}, \bibinfo {author} {\bibfnamefont {W.~F.}\ \bibnamefont
  {Reinhart}}, \bibinfo {author} {\bibfnamefont {T.}~\bibnamefont {Sanyal}},
  \bibinfo {author} {\bibfnamefont {M.~S.}\ \bibnamefont {Shell}}, \bibinfo
  {author} {\bibfnamefont {A.}~\bibnamefont {Nikoubashman}},\ and\ \bibinfo
  {author} {\bibfnamefont {A.~Z.}\ \bibnamefont {Panagiotopoulos}},\ }\bibfield
   {title} {\enquote {\bibinfo {title} {{Evaporation-induced assembly of
  colloidal crystals}},}\ }\href {https://doi.org/10.1063/1.5043401} {\bibfield
   {journal} {\bibinfo  {journal} {J. Chem. Phys.}\ }\textbf {\bibinfo {volume}
  {149}},\ \bibinfo {pages} {094901} (\bibinfo {year} {2018})}\BibitemShut
  {NoStop}%
\bibitem [{\citenamefont {Howard}\ and\ \citenamefont
  {Nikoubashman}(2020)}]{howard:jcp:2020}%
  \BibitemOpen
  \bibfield  {author} {\bibinfo {author} {\bibfnamefont {M.~P.}\ \bibnamefont
  {Howard}}\ and\ \bibinfo {author} {\bibfnamefont {A.}~\bibnamefont
  {Nikoubashman}},\ }\bibfield  {title} {\enquote {\bibinfo {title}
  {{Stratification of polymer mixtures in drying droplets: Hydrodynamics and
  diffusion}},}\ }\href {https://doi.org/10.1063/5.0014429} {\bibfield
  {journal} {\bibinfo  {journal} {J. Chem. Phys.}\ }\textbf {\bibinfo {volume}
  {153}},\ \bibinfo {pages} {054901} (\bibinfo {year} {2020})}\BibitemShut
  {NoStop}%
\bibitem [{\citenamefont {Yetkin}\ \emph {et~al.}(2024)\citenamefont {Yetkin},
  \citenamefont {Wani}, \citenamefont {Kritika}, \citenamefont {Howard},
  \citenamefont {Kappl}, \citenamefont {Butt},\ and\ \citenamefont
  {Nikoubashman}}]{yetkin:2024}%
  \BibitemOpen
  \bibfield  {author} {\bibinfo {author} {\bibfnamefont {M.}~\bibnamefont
  {Yetkin}}, \bibinfo {author} {\bibfnamefont {Y.~M.}\ \bibnamefont {Wani}},
  \bibinfo {author} {\bibfnamefont {K.}~\bibnamefont {Kritika}}, \bibinfo
  {author} {\bibfnamefont {M.~P.}\ \bibnamefont {Howard}}, \bibinfo {author}
  {\bibfnamefont {M.}~\bibnamefont {Kappl}}, \bibinfo {author} {\bibfnamefont
  {H.-J.}\ \bibnamefont {Butt}},\ and\ \bibinfo {author} {\bibfnamefont
  {A.}~\bibnamefont {Nikoubashman}},\ }\bibfield  {title} {\enquote {\bibinfo
  {title} {Structure formation in supraparticles composed of spherical and
  elongated particles},}\ }\href {https://doi.org/10.1021/acs.langmuir.3c03410}
  {\bibfield  {journal} {\bibinfo  {journal} {Langmuir}\ }\textbf {\bibinfo
  {volume} {40}},\ \bibinfo {pages} {1096--1108} (\bibinfo {year}
  {2024})}\BibitemShut {NoStop}%
\bibitem [{\citenamefont {Marconi}\ and\ \citenamefont
  {Tarazona}(1999)}]{marconi:jcp:1999}%
  \BibitemOpen
  \bibfield  {author} {\bibinfo {author} {\bibfnamefont {U.~M.~B.}\
  \bibnamefont {Marconi}}\ and\ \bibinfo {author} {\bibfnamefont
  {P.}~\bibnamefont {Tarazona}},\ }\bibfield  {title} {\enquote {\bibinfo
  {title} {Dynamic density functional theory of fluids},}\ }\href
  {https://doi.org/10.1063/1.478705} {\bibfield  {journal} {\bibinfo  {journal}
  {J. Chem. Phys.}\ }\textbf {\bibinfo {volume} {110}},\ \bibinfo {pages}
  {8032--8044} (\bibinfo {year} {1999})}\BibitemShut {NoStop}%
\bibitem [{\citenamefont {Rex}\ and\ \citenamefont
  {L\"{o}wen}(2008)}]{rex:prl:2008}%
  \BibitemOpen
  \bibfield  {author} {\bibinfo {author} {\bibfnamefont {M.}~\bibnamefont
  {Rex}}\ and\ \bibinfo {author} {\bibfnamefont {H.}~\bibnamefont
  {L\"{o}wen}},\ }\bibfield  {title} {\enquote {\bibinfo {title} {Dynamical
  density functional theory with hydrodynamic interactions and colloids in
  unstable traps},}\ }\href {https://doi.org/10.1103/PhysRevLett.101.148302}
  {\bibfield  {journal} {\bibinfo  {journal} {Phys. Rev. Lett.}\ }\textbf
  {\bibinfo {volume} {101}},\ \bibinfo {pages} {148302} (\bibinfo {year}
  {2008})}\BibitemShut {NoStop}%
\bibitem [{\citenamefont {Rex}\ and\ \citenamefont
  {L\"{o}wen}(2009)}]{rex:epje:2009}%
  \BibitemOpen
  \bibfield  {author} {\bibinfo {author} {\bibfnamefont {M.}~\bibnamefont
  {Rex}}\ and\ \bibinfo {author} {\bibfnamefont {H.}~\bibnamefont
  {L\"{o}wen}},\ }\bibfield  {title} {\enquote {\bibinfo {title} {Dynamical
  density functional theory for colloidal dispersions including hydrodynamic
  interactions},}\ }\href {https://doi.org/10.1140/epje/i2008-10363-x}
  {\bibfield  {journal} {\bibinfo  {journal} {Eur. Phys. J. E}\ }\textbf
  {\bibinfo {volume} {28}},\ \bibinfo {pages} {139--146} (\bibinfo {year}
  {2009})}\BibitemShut {NoStop}%
\bibitem [{\citenamefont {Howard}, \citenamefont {Nikoubashman},\ and\
  \citenamefont {Panagiotopoulos}(2017{\natexlab{b}})}]{howard:lng:2017B}%
  \BibitemOpen
  \bibfield  {author} {\bibinfo {author} {\bibfnamefont {M.~P.}\ \bibnamefont
  {Howard}}, \bibinfo {author} {\bibfnamefont {A.}~\bibnamefont
  {Nikoubashman}},\ and\ \bibinfo {author} {\bibfnamefont {A.~Z.}\ \bibnamefont
  {Panagiotopoulos}},\ }\bibfield  {title} {\enquote {\bibinfo {title}
  {Stratification in drying polymer-polymer and colloid-polymer mixtures},}\
  }\href@noop {} {\bibfield  {journal} {\bibinfo  {journal} {Langmuir}\
  }\textbf {\bibinfo {volume} {33}},\ \bibinfo {pages} {11390--11398} (\bibinfo
  {year} {2017}{\natexlab{b}})}\BibitemShut {NoStop}%
\bibitem [{\citenamefont {Langmuir.}(1918)}]{langmuir:1918}%
  \BibitemOpen
  \bibfield  {author} {\bibinfo {author} {\bibfnamefont {I.}~\bibnamefont
  {Langmuir.}},\ }\bibfield  {title} {\enquote {\bibinfo {title} {The
  evaporation of small spheres},}\ }\href
  {https://doi.org/10.1103/PhysRev.12.368} {\bibfield  {journal} {\bibinfo
  {journal} {Phys. Rev.}\ }\textbf {\bibinfo {volume} {12}},\ \bibinfo {pages}
  {368--370} (\bibinfo {year} {1918})}\BibitemShut {NoStop}%
\bibitem [{\citenamefont {Roth}(2010)}]{Roth:2010ei}%
  \BibitemOpen
  \bibfield  {author} {\bibinfo {author} {\bibfnamefont {R.}~\bibnamefont
  {Roth}},\ }\bibfield  {title} {\enquote {\bibinfo {title} {Fundamental
  measure theory for hard-sphere mixtures: a review},}\ }\href@noop {}
  {\bibfield  {journal} {\bibinfo  {journal} {J. Phys.: Condens. Matter}\
  }\textbf {\bibinfo {volume} {22}},\ \bibinfo {pages} {063102} (\bibinfo
  {year} {2010})}\BibitemShut {NoStop}%
\bibitem [{\citenamefont {Reinhardt}\ and\ \citenamefont
  {Brader}(2012)}]{reinhardt:2012}%
  \BibitemOpen
  \bibfield  {author} {\bibinfo {author} {\bibfnamefont {J.}~\bibnamefont
  {Reinhardt}}\ and\ \bibinfo {author} {\bibfnamefont {J.~M.}\ \bibnamefont
  {Brader}},\ }\bibfield  {title} {\enquote {\bibinfo {title} {Dynamics of
  localized particles from density functional theory},}\ }\href
  {https://doi.org/10.1103/PhysRevE.85.011404} {\bibfield  {journal} {\bibinfo
  {journal} {Phys. Rev. E}\ }\textbf {\bibinfo {volume} {85}},\ \bibinfo
  {pages} {011404} (\bibinfo {year} {2012})}\BibitemShut {NoStop}%
\bibitem [{\citenamefont {Schulz}\ and\ \citenamefont
  {Keddie}(2018)}]{schulz:sm:2018}%
  \BibitemOpen
  \bibfield  {author} {\bibinfo {author} {\bibfnamefont {M.}~\bibnamefont
  {Schulz}}\ and\ \bibinfo {author} {\bibfnamefont {J.}~\bibnamefont
  {Keddie}},\ }\bibfield  {title} {\enquote {\bibinfo {title} {A critical and
  quantitative review of the stratification of particles during the drying of
  colloidal films},}\ }\href@noop {} {\bibfield  {journal} {\bibinfo  {journal}
  {Soft Matter}\ }\textbf {\bibinfo {volume} {14}},\ \bibinfo {pages}
  {6181--6197} (\bibinfo {year} {2018})}\BibitemShut {NoStop}%
\bibitem [{\citenamefont {Bahadur}\ \emph {et~al.}(2011)\citenamefont
  {Bahadur}, \citenamefont {Sen}, \citenamefont {Mazumder}, \citenamefont
  {Bhattacharya}, \citenamefont {Frielinghaus},\ and\ \citenamefont
  {Goerigk}}]{bahadur:2011}%
  \BibitemOpen
  \bibfield  {author} {\bibinfo {author} {\bibfnamefont {J.}~\bibnamefont
  {Bahadur}}, \bibinfo {author} {\bibfnamefont {D.}~\bibnamefont {Sen}},
  \bibinfo {author} {\bibfnamefont {S.}~\bibnamefont {Mazumder}}, \bibinfo
  {author} {\bibfnamefont {S.}~\bibnamefont {Bhattacharya}}, \bibinfo {author}
  {\bibfnamefont {H.}~\bibnamefont {Frielinghaus}},\ and\ \bibinfo {author}
  {\bibfnamefont {G.}~\bibnamefont {Goerigk}},\ }\bibfield  {title} {\enquote
  {\bibinfo {title} {Origin of buckling phenomenon during drying of
  micrometer-sized colloidal droplets},}\ }\href
  {https://doi.org/10.1021/la200827n} {\bibfield  {journal} {\bibinfo
  {journal} {Langmuir}\ }\textbf {\bibinfo {volume} {27}},\ \bibinfo {pages}
  {8404--8414} (\bibinfo {year} {2011})}\BibitemShut {NoStop}%
\bibitem [{\citenamefont {Weeks}, \citenamefont {Chandler},\ and\ \citenamefont
  {Andersen}(1971)}]{weeks:1971}%
  \BibitemOpen
  \bibfield  {author} {\bibinfo {author} {\bibfnamefont {J.~D.}\ \bibnamefont
  {Weeks}}, \bibinfo {author} {\bibfnamefont {D.}~\bibnamefont {Chandler}},\
  and\ \bibinfo {author} {\bibfnamefont {H.~C.}\ \bibnamefont {Andersen}},\
  }\bibfield  {title} {\enquote {\bibinfo {title} {Role of repulsive forces in
  determining equilibrium structure of simple liquids},}\ }\href
  {https://doi.org/10.1063/1.1674820} {\bibfield  {journal} {\bibinfo
  {journal} {J. Chem. Phys.}\ }\textbf {\bibinfo {volume} {54}},\ \bibinfo
  {pages} {5237--5247} (\bibinfo {year} {1971})}\BibitemShut {NoStop}%
\bibitem [{\citenamefont {Batchelor}\ and\ \citenamefont
  {Green}(1972)}]{batchelor_green:jfm:1972}%
  \BibitemOpen
  \bibfield  {author} {\bibinfo {author} {\bibfnamefont {G.~K.}\ \bibnamefont
  {Batchelor}}\ and\ \bibinfo {author} {\bibfnamefont {J.~T.}\ \bibnamefont
  {Green}},\ }\bibfield  {title} {\enquote {\bibinfo {title} {The hydrodynamic
  interaction of two small freely-moving spheres in a linear flow field},}\
  }\href {https://doi.org/10.1017/S0022112072002927} {\bibfield  {journal}
  {\bibinfo  {journal} {J. Fluid Mech.}\ }\textbf {\bibinfo {volume} {56}},\
  \bibinfo {pages} {375–400} (\bibinfo {year} {1972})}\BibitemShut {NoStop}%
\bibitem [{\citenamefont {Batchelor}(1983)}]{batchelor:jfm:1983}%
  \BibitemOpen
  \bibfield  {author} {\bibinfo {author} {\bibfnamefont {G.~K.}\ \bibnamefont
  {Batchelor}},\ }\bibfield  {title} {\enquote {\bibinfo {title} {Diffusion in
  a dilute polydisperse system of interacting spheres},}\ }\href@noop {}
  {\bibfield  {journal} {\bibinfo  {journal} {J. Fluid Mech.}\ }\textbf
  {\bibinfo {volume} {131}},\ \bibinfo {pages} {155--175} (\bibinfo {year}
  {1983})}\BibitemShut {NoStop}%
\bibitem [{\citenamefont {Brady}\ and\ \citenamefont
  {Bossis}(1988)}]{brady:arfm:1988}%
  \BibitemOpen
  \bibfield  {author} {\bibinfo {author} {\bibfnamefont {J.~F.}\ \bibnamefont
  {Brady}}\ and\ \bibinfo {author} {\bibfnamefont {G.}~\bibnamefont {Bossis}},\
  }\bibfield  {title} {\enquote {\bibinfo {title} {Stokesian dynamics},}\
  }\href {https://doi.org/https://doi.org/10.1146/annurev.fl.20.010188.000551}
  {\bibfield  {journal} {\bibinfo  {journal} {Annu. Rev. Fluid Mech.}\ }\textbf
  {\bibinfo {volume} {20}},\ \bibinfo {pages} {111--157} (\bibinfo {year}
  {1988})}\BibitemShut {NoStop}%
\bibitem [{\citenamefont {Russel}, \citenamefont {Saville},\ and\ \citenamefont
  {Schowalter}(1989)}]{russel:1989}%
  \BibitemOpen
  \bibfield  {author} {\bibinfo {author} {\bibfnamefont {W.~B.}\ \bibnamefont
  {Russel}}, \bibinfo {author} {\bibfnamefont {D.~A.}\ \bibnamefont
  {Saville}},\ and\ \bibinfo {author} {\bibfnamefont {W.~R.}\ \bibnamefont
  {Schowalter}},\ }\href {https://doi.org/10.1017/CBO9780511608810} {\emph
  {\bibinfo {title} {Colloidal dispersions}}}\ (\bibinfo  {publisher}
  {Cambridge University Press},\ \bibinfo {address} {New York},\ \bibinfo
  {year} {1989})\BibitemShut {NoStop}%
\bibitem [{\citenamefont {Happel}\ and\ \citenamefont
  {Brenner}(2012)}]{happel:2012}%
  \BibitemOpen
  \bibfield  {author} {\bibinfo {author} {\bibfnamefont {J.}~\bibnamefont
  {Happel}}\ and\ \bibinfo {author} {\bibfnamefont {H.}~\bibnamefont
  {Brenner}},\ }\href@noop {} {\emph {\bibinfo {title} {Low Reynolds number
  hydrodynamics: with special applications to particulate media}}},\
  Vol.~\bibinfo {volume} {1}\ (\bibinfo  {publisher} {Springer Science \&
  Business Media},\ \bibinfo {year} {2012})\BibitemShut {NoStop}%
\bibitem [{\citenamefont {Rotne}\ and\ \citenamefont
  {Prager}(1969)}]{rotne:1969}%
  \BibitemOpen
  \bibfield  {author} {\bibinfo {author} {\bibfnamefont {J.}~\bibnamefont
  {Rotne}}\ and\ \bibinfo {author} {\bibfnamefont {S.}~\bibnamefont {Prager}},\
  }\bibfield  {title} {\enquote {\bibinfo {title} {Variational treatment of
  hydrodynamic interaction in polymers},}\ }\href
  {https://doi.org/10.1063/1.1670977} {\bibfield  {journal} {\bibinfo
  {journal} {J. Chem. Phys.}\ }\textbf {\bibinfo {volume} {50}},\ \bibinfo
  {pages} {4831--4837} (\bibinfo {year} {1969})}\BibitemShut {NoStop}%
\bibitem [{\citenamefont {Yamakawa}(1970)}]{yamakawa:1970}%
  \BibitemOpen
  \bibfield  {author} {\bibinfo {author} {\bibfnamefont {H.}~\bibnamefont
  {Yamakawa}},\ }\bibfield  {title} {\enquote {\bibinfo {title} {Transport
  properties of polymer chains in dilute solution: Hydrodynamic interaction},}\
  }\href {https://doi.org/10.1063/1.1673799} {\bibfield  {journal} {\bibinfo
  {journal} {J. Chem. Phys.}\ }\textbf {\bibinfo {volume} {53}},\ \bibinfo
  {pages} {436--443} (\bibinfo {year} {1970})}\BibitemShut {NoStop}%
\bibitem [{\citenamefont {Anderson}, \citenamefont {Glaser},\ and\
  \citenamefont {Glotzer}(2020)}]{anderson:cms:2020}%
  \BibitemOpen
  \bibfield  {author} {\bibinfo {author} {\bibfnamefont {J.~A.}\ \bibnamefont
  {Anderson}}, \bibinfo {author} {\bibfnamefont {J.}~\bibnamefont {Glaser}},\
  and\ \bibinfo {author} {\bibfnamefont {S.~C.}\ \bibnamefont {Glotzer}},\
  }\bibfield  {title} {\enquote {\bibinfo {title} {Hoomd-blue: A python package
  for high-performance molecular dynamics and hard particle monte carlo
  simulations},}\ }\href@noop {} {\bibfield  {journal} {\bibinfo  {journal}
  {Comput. Mater. Sci.}\ }\textbf {\bibinfo {volume} {173}},\ \bibinfo {pages}
  {109363} (\bibinfo {year} {2020})}\BibitemShut {NoStop}%
\bibitem [{azp()}]{azplugins}%
  \BibitemOpen
  \href@noop {} {}\bibinfo {howpublished}
  {https://github.com/mphowardlab/azplugins}\BibitemShut {NoStop}%
\bibitem [{\citenamefont {Ermak}\ and\ \citenamefont
  {McCammon}(1978)}]{ermak:jcp:1978}%
  \BibitemOpen
  \bibfield  {author} {\bibinfo {author} {\bibfnamefont {D.~J.}\ \bibnamefont
  {Ermak}}\ and\ \bibinfo {author} {\bibfnamefont {J.~A.}\ \bibnamefont
  {McCammon}},\ }\bibfield  {title} {\enquote {\bibinfo {title} {Brownian
  dynamics with hydrodynamic interactions},}\ }\href@noop {} {\bibfield
  {journal} {\bibinfo  {journal} {J. Chem. Phys.}\ }\textbf {\bibinfo {volume}
  {69}},\ \bibinfo {pages} {1352--1360} (\bibinfo {year} {1978})}\BibitemShut
  {NoStop}%
\bibitem [{\citenamefont {Beenakker}(1986)}]{beenakker:1986}%
  \BibitemOpen
  \bibfield  {author} {\bibinfo {author} {\bibfnamefont {C.}~\bibnamefont
  {Beenakker}},\ }\bibfield  {title} {\enquote {\bibinfo {title} {Ewald sum of
  the rotne--prager tensor},}\ }\href@noop {} {\bibfield  {journal} {\bibinfo
  {journal} {J. Chem. Phys.}\ }\textbf {\bibinfo {volume} {85}},\ \bibinfo
  {pages} {1581--1582} (\bibinfo {year} {1986})}\BibitemShut {NoStop}%
\bibitem [{\citenamefont {Fiore}\ \emph {et~al.}(2017)\citenamefont {Fiore},
  \citenamefont {Balboa~Usabiaga}, \citenamefont {Donev},\ and\ \citenamefont
  {Swan}}]{fiore:2017}%
  \BibitemOpen
  \bibfield  {author} {\bibinfo {author} {\bibfnamefont {A.~M.}\ \bibnamefont
  {Fiore}}, \bibinfo {author} {\bibfnamefont {F.}~\bibnamefont
  {Balboa~Usabiaga}}, \bibinfo {author} {\bibfnamefont {A.}~\bibnamefont
  {Donev}},\ and\ \bibinfo {author} {\bibfnamefont {J.~W.}\ \bibnamefont
  {Swan}},\ }\bibfield  {title} {\enquote {\bibinfo {title} {{Rapid sampling of
  stochastic displacements in Brownian dynamics simulations}},}\ }\href
  {https://doi.org/10.1063/1.4978242} {\bibfield  {journal} {\bibinfo
  {journal} {J. Chem. Phys.}\ }\textbf {\bibinfo {volume} {146}},\ \bibinfo
  {pages} {124116} (\bibinfo {year} {2017})}\BibitemShut {NoStop}%
\bibitem [{\citenamefont {Aponte-Rivera}\ and\ \citenamefont
  {Zia}(2016)}]{aponte-rivera:2016}%
  \BibitemOpen
  \bibfield  {author} {\bibinfo {author} {\bibfnamefont {C.}~\bibnamefont
  {Aponte-Rivera}}\ and\ \bibinfo {author} {\bibfnamefont {R.~N.}\ \bibnamefont
  {Zia}},\ }\bibfield  {title} {\enquote {\bibinfo {title} {Simulation of
  hydrodynamically interacting particles confined by a spherical cavity},}\
  }\href {https://doi.org/10.1103/PhysRevFluids.1.023301} {\bibfield  {journal}
  {\bibinfo  {journal} {Phys. Rev. Fluids}\ }\textbf {\bibinfo {volume} {1}},\
  \bibinfo {pages} {023301} (\bibinfo {year} {2016})}\BibitemShut {NoStop}%
\bibitem [{\citenamefont {Howard}, \citenamefont {Panagiotopoulos},\ and\
  \citenamefont {Nikoubashman}(2018{\natexlab{a}})}]{howard:cpc:2018}%
  \BibitemOpen
  \bibfield  {author} {\bibinfo {author} {\bibfnamefont {M.~P.}\ \bibnamefont
  {Howard}}, \bibinfo {author} {\bibfnamefont {A.~Z.}\ \bibnamefont
  {Panagiotopoulos}},\ and\ \bibinfo {author} {\bibfnamefont {A.}~\bibnamefont
  {Nikoubashman}},\ }\bibfield  {title} {\enquote {\bibinfo {title} {Efficient
  mesoscale hydrodynamics: Multiparticle collision dynamics with massively
  parallel gpu acceleration},}\ }\href@noop {} {\bibfield  {journal} {\bibinfo
  {journal} {Comput. Phys. Commun.}\ }\textbf {\bibinfo {volume} {230}},\
  \bibinfo {pages} {10--20} (\bibinfo {year} {2018}{\natexlab{a}})}\BibitemShut
  {NoStop}%
\bibitem [{\citenamefont {Ihle}\ and\ \citenamefont {Kroll}(2001)}]{ihle:2001}%
  \BibitemOpen
  \bibfield  {author} {\bibinfo {author} {\bibfnamefont {T.}~\bibnamefont
  {Ihle}}\ and\ \bibinfo {author} {\bibfnamefont {D.~M.}\ \bibnamefont
  {Kroll}},\ }\bibfield  {title} {\enquote {\bibinfo {title} {Stochastic
  rotation dynamics: A galilean-invariant mesoscopic model for fluid flow},}\
  }\href {https://doi.org/10.1103/PhysRevE.63.020201} {\bibfield  {journal}
  {\bibinfo  {journal} {Phys. Rev. E}\ }\textbf {\bibinfo {volume} {63}},\
  \bibinfo {pages} {020201} (\bibinfo {year} {2001})}\BibitemShut {NoStop}%
\bibitem [{\citenamefont {Gompper}\ \emph {et~al.}(2008)\citenamefont
  {Gompper}, \citenamefont {Ihle}, \citenamefont {Kroll},\ and\ \citenamefont
  {Winkler}}]{gompper:2008}%
  \BibitemOpen
  \bibfield  {author} {\bibinfo {author} {\bibfnamefont {G.}~\bibnamefont
  {Gompper}}, \bibinfo {author} {\bibfnamefont {T.}~\bibnamefont {Ihle}},
  \bibinfo {author} {\bibfnamefont {D.}~\bibnamefont {Kroll}},\ and\ \bibinfo
  {author} {\bibfnamefont {R.}~\bibnamefont {Winkler}},\ }\bibfield  {title}
  {\enquote {\bibinfo {title} {Multi-particle collision dynamics: A
  particle-based mesoscale simulation approach to the hydrodynamics of complex
  fluids},}\ }\href {https://doi.org/10.1007/978-3-540-87706-6_1} {\bibfield
  {journal} {\bibinfo  {journal} {Adv. Polym. Sci.}\ }\textbf {\bibinfo
  {volume} {221}},\ \bibinfo {pages} {1--87} (\bibinfo {year}
  {2008})}\BibitemShut {NoStop}%
\bibitem [{\citenamefont {Wani}\ \emph {et~al.}(2022)\citenamefont {Wani},
  \citenamefont {Kovakas}, \citenamefont {Nikoubashman},\ and\ \citenamefont
  {Howard}}]{wani:2022}%
  \BibitemOpen
  \bibfield  {author} {\bibinfo {author} {\bibfnamefont {Y.~M.}\ \bibnamefont
  {Wani}}, \bibinfo {author} {\bibfnamefont {P.~G.}\ \bibnamefont {Kovakas}},
  \bibinfo {author} {\bibfnamefont {A.}~\bibnamefont {Nikoubashman}},\ and\
  \bibinfo {author} {\bibfnamefont {M.~P.}\ \bibnamefont {Howard}},\ }\bibfield
   {title} {\enquote {\bibinfo {title} {{Diffusion and sedimentation in
  colloidal suspensions using multiparticle collision dynamics with a discrete
  particle model}},}\ }\href {https://doi.org/10.1063/5.0075002} {\bibfield
  {journal} {\bibinfo  {journal} {J. Chem. Phys.}\ }\textbf {\bibinfo {volume}
  {156}},\ \bibinfo {pages} {024901} (\bibinfo {year} {2022})}\BibitemShut
  {NoStop}%
\bibitem [{\citenamefont {Wani}\ \emph {et~al.}(2024)\citenamefont {Wani},
  \citenamefont {Kovakas}, \citenamefont {Nikoubashman},\ and\ \citenamefont
  {Howard}}]{wani:2024}%
  \BibitemOpen
  \bibfield  {author} {\bibinfo {author} {\bibfnamefont {Y.~M.}\ \bibnamefont
  {Wani}}, \bibinfo {author} {\bibfnamefont {P.~G.}\ \bibnamefont {Kovakas}},
  \bibinfo {author} {\bibfnamefont {A.}~\bibnamefont {Nikoubashman}},\ and\
  \bibinfo {author} {\bibfnamefont {M.~P.}\ \bibnamefont {Howard}},\ }\bibfield
   {title} {\enquote {\bibinfo {title} {Mesoscale simulations of diffusion and
  sedimentation in shape-anisotropic nanoparticle suspensions},}\ }\href
  {https://doi.org/10.1039/D4SM00271G} {\bibfield  {journal} {\bibinfo
  {journal} {Soft Matter}\ }\textbf {\bibinfo {volume} {20}},\ \bibinfo {pages}
  {3942--3953} (\bibinfo {year} {2024})}\BibitemShut {NoStop}%
\bibitem [{\citenamefont {Huang}\ \emph {et~al.}(2010)\citenamefont {Huang},
  \citenamefont {Chatterji}, \citenamefont {Sutmann}, \citenamefont {Gompper},\
  and\ \citenamefont {Winkler}}]{haung:2010}%
  \BibitemOpen
  \bibfield  {author} {\bibinfo {author} {\bibfnamefont {C.}~\bibnamefont
  {Huang}}, \bibinfo {author} {\bibfnamefont {A.}~\bibnamefont {Chatterji}},
  \bibinfo {author} {\bibfnamefont {G.}~\bibnamefont {Sutmann}}, \bibinfo
  {author} {\bibfnamefont {G.}~\bibnamefont {Gompper}},\ and\ \bibinfo {author}
  {\bibfnamefont {R.}~\bibnamefont {Winkler}},\ }\bibfield  {title} {\enquote
  {\bibinfo {title} {Cell-level canonical sampling by velocity scaling for
  multiparticle collision dynamics simulations},}\ }\href
  {https://doi.org/https://doi.org/10.1016/j.jcp.2009.09.024} {\bibfield
  {journal} {\bibinfo  {journal} {J. Comput. Phys.}\ }\textbf {\bibinfo
  {volume} {229}},\ \bibinfo {pages} {168--177} (\bibinfo {year}
  {2010})}\BibitemShut {NoStop}%
\bibitem [{\citenamefont {Poblete}\ \emph {et~al.}(2014)\citenamefont
  {Poblete}, \citenamefont {Wysocki}, \citenamefont {Gompper},\ and\
  \citenamefont {Winkler}}]{poblete:2014}%
  \BibitemOpen
  \bibfield  {author} {\bibinfo {author} {\bibfnamefont {S.}~\bibnamefont
  {Poblete}}, \bibinfo {author} {\bibfnamefont {A.}~\bibnamefont {Wysocki}},
  \bibinfo {author} {\bibfnamefont {G.}~\bibnamefont {Gompper}},\ and\ \bibinfo
  {author} {\bibfnamefont {R.~G.}\ \bibnamefont {Winkler}},\ }\bibfield
  {title} {\enquote {\bibinfo {title} {Hydrodynamics of discrete-particle
  models of spherical colloids: A multiparticle collision dynamics simulation
  study},}\ }\href {https://doi.org/10.1103/PhysRevE.90.033314} {\bibfield
  {journal} {\bibinfo  {journal} {Phys. Rev. E}\ }\textbf {\bibinfo {volume}
  {90}},\ \bibinfo {pages} {033314} (\bibinfo {year} {2014})}\BibitemShut
  {NoStop}%
\bibitem [{\citenamefont {Peng}\ and\ \citenamefont {Sinno}(2024)}]{peng:2024}%
  \BibitemOpen
  \bibfield  {author} {\bibinfo {author} {\bibfnamefont {Y.-S.}\ \bibnamefont
  {Peng}}\ and\ \bibinfo {author} {\bibfnamefont {T.}~\bibnamefont {Sinno}},\
  }\bibfield  {title} {\enquote {\bibinfo {title} {{Multiparticle collision
  dynamics simulations of hydrodynamic interactions in colloidal suspensions:
  How well does the discrete particle approach do at short range?}}}\ }\href
  {https://doi.org/10.1063/5.0197818} {\bibfield  {journal} {\bibinfo
  {journal} {J. Chem. Phys.}\ }\textbf {\bibinfo {volume} {160}},\ \bibinfo
  {pages} {174121} (\bibinfo {year} {2024})}\BibitemShut {NoStop}%
\bibitem [{\citenamefont {Howard}, \citenamefont {Panagiotopoulos},\ and\
  \citenamefont {Nikoubashman}(2018{\natexlab{b}})}]{howard:2018}%
  \BibitemOpen
  \bibfield  {author} {\bibinfo {author} {\bibfnamefont {M.~P.}\ \bibnamefont
  {Howard}}, \bibinfo {author} {\bibfnamefont {A.~Z.}\ \bibnamefont
  {Panagiotopoulos}},\ and\ \bibinfo {author} {\bibfnamefont {A.}~\bibnamefont
  {Nikoubashman}},\ }\bibfield  {title} {\enquote {\bibinfo {title} {Efficient
  mesoscale hydrodynamics: Multiparticle collision dynamics with massively
  parallel gpu acceleration},}\ }\href
  {https://doi.org/https://doi.org/10.1016/j.cpc.2018.04.009} {\bibfield
  {journal} {\bibinfo  {journal} {Comput. Phys. Commun.}\ }\textbf {\bibinfo
  {volume} {230}},\ \bibinfo {pages} {10--20} (\bibinfo {year}
  {2018}{\natexlab{b}})}\BibitemShut {NoStop}%
\bibitem [{\citenamefont {Archer}\ and\ \citenamefont
  {Evans}(2004)}]{archer_evans:2004}%
  \BibitemOpen
  \bibfield  {author} {\bibinfo {author} {\bibfnamefont {A.~J.}\ \bibnamefont
  {Archer}}\ and\ \bibinfo {author} {\bibfnamefont {R.}~\bibnamefont {Evans}},\
  }\bibfield  {title} {\enquote {\bibinfo {title} {{Dynamical density
  functional theory and its application to spinodal decomposition}},}\ }\href
  {https://doi.org/10.1063/1.1778374} {\bibfield  {journal} {\bibinfo
  {journal} {J. Chem. Phys.}\ }\textbf {\bibinfo {volume} {121}},\ \bibinfo
  {pages} {4246--4254} (\bibinfo {year} {2004})}\BibitemShut {NoStop}%
\bibitem [{\citenamefont {Archer}(2005)}]{archer:jpcm:2005}%
  \BibitemOpen
  \bibfield  {author} {\bibinfo {author} {\bibfnamefont {A.~J.}\ \bibnamefont
  {Archer}},\ }\bibfield  {title} {\enquote {\bibinfo {title} {Dynamical
  density functional theory: binary phase-separating colloidal fluid in a
  cavity},}\ }\href {https://doi.org/10.1088/0953-8984/17/10/001} {\bibfield
  {journal} {\bibinfo  {journal} {J. Phys.: Condens. Matter}\ }\textbf
  {\bibinfo {volume} {17}},\ \bibinfo {pages} {1405--1427} (\bibinfo {year}
  {2005})}\BibitemShut {NoStop}%
\bibitem [{\citenamefont {Archer}(2009)}]{archer:2009}%
  \BibitemOpen
  \bibfield  {author} {\bibinfo {author} {\bibfnamefont {A.~J.}\ \bibnamefont
  {Archer}},\ }\bibfield  {title} {\enquote {\bibinfo {title} {{Dynamical
  density functional theory for molecular and colloidal fluids: A microscopic
  approach to fluid mechanics}},}\ }\href {https://doi.org/10.1063/1.3054633}
  {\bibfield  {journal} {\bibinfo  {journal} {J. Chem. Phys.}\ }\textbf
  {\bibinfo {volume} {130}},\ \bibinfo {pages} {014509} (\bibinfo {year}
  {2009})}\BibitemShut {NoStop}%
\bibitem [{\citenamefont {Rosenfeld}(1989)}]{Rosenfeld:1989uh}%
  \BibitemOpen
  \bibfield  {author} {\bibinfo {author} {\bibfnamefont {Y.}~\bibnamefont
  {Rosenfeld}},\ }\bibfield  {title} {\enquote {\bibinfo {title} {Free-energy
  model for the inhomogeneous hard-sphere fluid mixture and density-functional
  theory of freezing},}\ }\href@noop {} {\bibfield  {journal} {\bibinfo
  {journal} {Phys. Rev. Lett.}\ }\textbf {\bibinfo {volume} {63}},\ \bibinfo
  {pages} {980--983} (\bibinfo {year} {1989})}\BibitemShut {NoStop}%
\bibitem [{\citenamefont {te~Vrugt}, \citenamefont {L\"{o}wen},\ and\
  \citenamefont {Wittkowski}(2020)}]{tevrugt:advphys:2020}%
  \BibitemOpen
  \bibfield  {author} {\bibinfo {author} {\bibfnamefont {M.}~\bibnamefont
  {te~Vrugt}}, \bibinfo {author} {\bibfnamefont {H.}~\bibnamefont
  {L\"{o}wen}},\ and\ \bibinfo {author} {\bibfnamefont {R.}~\bibnamefont
  {Wittkowski}},\ }\bibfield  {title} {\enquote {\bibinfo {title} {Classical
  dynamical density functional theory: From fundamentals to applications},}\
  }\href@noop {} {\bibfield  {journal} {\bibinfo  {journal} {Adv. Phys.}\
  }\textbf {\bibinfo {volume} {69}},\ \bibinfo {pages} {121--247} (\bibinfo
  {year} {2020})}\BibitemShut {NoStop}%
\bibitem [{\citenamefont {Hansen}\ and\ \citenamefont
  {McDonald}(2006)}]{hansen:2006}%
  \BibitemOpen
  \bibfield  {author} {\bibinfo {author} {\bibfnamefont {J.~P.}\ \bibnamefont
  {Hansen}}\ and\ \bibinfo {author} {\bibfnamefont {I.~R.}\ \bibnamefont
  {McDonald}},\ }\href@noop {} {\emph {\bibinfo {title} {Theory of Simple
  Liquids}}},\ \bibinfo {edition} {3rd}\ ed.\ (\bibinfo  {publisher} {Academic
  Press},\ \bibinfo {address} {Amsterdam},\ \bibinfo {year} {2006})\BibitemShut
  {NoStop}%
\bibitem [{\citenamefont {Goddard}\ \emph {et~al.}(2013)\citenamefont
  {Goddard}, \citenamefont {Nold}, \citenamefont {Savva}, \citenamefont
  {Yatsyshin},\ and\ \citenamefont {Kalliadasis}}]{goddard:2013}%
  \BibitemOpen
  \bibfield  {author} {\bibinfo {author} {\bibfnamefont {B.~D.}\ \bibnamefont
  {Goddard}}, \bibinfo {author} {\bibfnamefont {A.}~\bibnamefont {Nold}},
  \bibinfo {author} {\bibfnamefont {N.}~\bibnamefont {Savva}}, \bibinfo
  {author} {\bibfnamefont {P.}~\bibnamefont {Yatsyshin}},\ and\ \bibinfo
  {author} {\bibfnamefont {S.}~\bibnamefont {Kalliadasis}},\ }\bibfield
  {title} {\enquote {\bibinfo {title} {Unification of dynamic density
  functional theory for colloidal fluids to include inertia and hydrodynamic
  interactions: derivation and numerical experiments},}\ }\href
  {https://doi.org/10.1088/0953-8984/25/3/035101} {\bibfield  {journal}
  {\bibinfo  {journal} {J. Phys.: Condens. Matter}\ }\textbf {\bibinfo {volume}
  {25}},\ \bibinfo {pages} {035101} (\bibinfo {year} {2013})}\BibitemShut
  {NoStop}%
\bibitem [{\citenamefont {Donev}\ and\ \citenamefont
  {Vanden-Eijnden}(2014)}]{donev:2014}%
  \BibitemOpen
  \bibfield  {author} {\bibinfo {author} {\bibfnamefont {A.}~\bibnamefont
  {Donev}}\ and\ \bibinfo {author} {\bibfnamefont {E.}~\bibnamefont
  {Vanden-Eijnden}},\ }\bibfield  {title} {\enquote {\bibinfo {title} {Dynamic
  density functional theory with hydrodynamic interactions and fluctuations},}\
  }\href {https://doi.org/10.1063/1.4883520} {\bibfield  {journal} {\bibinfo
  {journal} {J. Chem. Phys.}\ }\textbf {\bibinfo {volume} {140}},\ \bibinfo
  {pages} {234115} (\bibinfo {year} {2014})}\BibitemShut {NoStop}%
\bibitem [{\citenamefont {Goddard}, \citenamefont {Nold},\ and\ \citenamefont
  {Kalliadasis}(2016)}]{goddard:2016}%
  \BibitemOpen
  \bibfield  {author} {\bibinfo {author} {\bibfnamefont {B.~D.}\ \bibnamefont
  {Goddard}}, \bibinfo {author} {\bibfnamefont {A.}~\bibnamefont {Nold}},\ and\
  \bibinfo {author} {\bibfnamefont {S.}~\bibnamefont {Kalliadasis}},\
  }\bibfield  {title} {\enquote {\bibinfo {title} {{Dynamical density
  functional theory with hydrodynamic interactions in confined geometries}},}\
  }\href {https://doi.org/10.1063/1.4968565} {\bibfield  {journal} {\bibinfo
  {journal} {J. Chem. Phys.}\ }\textbf {\bibinfo {volume} {145}},\ \bibinfo
  {pages} {214106} (\bibinfo {year} {2016})}\BibitemShut {NoStop}%
\bibitem [{\citenamefont {Goddard}, \citenamefont {Mills-Williams},\ and\
  \citenamefont {Sun}(2020)}]{goddard:2020}%
  \BibitemOpen
  \bibfield  {author} {\bibinfo {author} {\bibfnamefont {B.~D.}\ \bibnamefont
  {Goddard}}, \bibinfo {author} {\bibfnamefont {R.~D.}\ \bibnamefont
  {Mills-Williams}},\ and\ \bibinfo {author} {\bibfnamefont {J.}~\bibnamefont
  {Sun}},\ }\bibfield  {title} {\enquote {\bibinfo {title} {The singular
  hydrodynamic interactions between two spheres in stokes flow},}\ }\href
  {https://doi.org/10.1063/5.0009053} {\bibfield  {journal} {\bibinfo
  {journal} {Phys. Fluids}\ }\textbf {\bibinfo {volume} {32}},\ \bibinfo
  {pages} {062001} (\bibinfo {year} {2020})}\BibitemShut {NoStop}%
\bibitem [{\citenamefont {Sears}\ and\ \citenamefont
  {Frink}(2003)}]{sear:jcp:2003}%
  \BibitemOpen
  \bibfield  {author} {\bibinfo {author} {\bibfnamefont {M.~P.}\ \bibnamefont
  {Sears}}\ and\ \bibinfo {author} {\bibfnamefont {L.~J.}\ \bibnamefont
  {Frink}},\ }\bibfield  {title} {\enquote {\bibinfo {title} {A new efficient
  method for density functional theory calculations of inhomogeneous fluids},}\
  }\href {https://doi.org/https://doi.org/10.1016/S0021-9991(03)00270-5}
  {\bibfield  {journal} {\bibinfo  {journal} {J. Comput. Phys.}\ }\textbf
  {\bibinfo {volume} {190}},\ \bibinfo {pages} {184--200} (\bibinfo {year}
  {2003})}\BibitemShut {NoStop}%
\bibitem [{\citenamefont {Kierlik}\ and\ \citenamefont
  {Rosinberg}(1991)}]{kierlik:1991}%
  \BibitemOpen
  \bibfield  {author} {\bibinfo {author} {\bibfnamefont {E.}~\bibnamefont
  {Kierlik}}\ and\ \bibinfo {author} {\bibfnamefont {M.~L.}\ \bibnamefont
  {Rosinberg}},\ }\bibfield  {title} {\enquote {\bibinfo {title}
  {Density-functional theory for inhomogeneous fluids: Adsorption of binary
  mixtures},}\ }\href {https://doi.org/10.1103/PhysRevA.44.5025} {\bibfield
  {journal} {\bibinfo  {journal} {Phys. Rev. A}\ }\textbf {\bibinfo {volume}
  {44}},\ \bibinfo {pages} {5025--5037} (\bibinfo {year} {1991})}\BibitemShut
  {NoStop}%
\bibitem [{\citenamefont {Snook}\ and\ \citenamefont
  {Henderson}(1978)}]{snook:1978}%
  \BibitemOpen
  \bibfield  {author} {\bibinfo {author} {\bibfnamefont {I.~K.}\ \bibnamefont
  {Snook}}\ and\ \bibinfo {author} {\bibfnamefont {D.}~\bibnamefont
  {Henderson}},\ }\bibfield  {title} {\enquote {\bibinfo {title} {Monte carlo
  study of a hard‐sphere fluid near a hard wall},}\ }\href
  {https://doi.org/10.1063/1.436036} {\bibfield  {journal} {\bibinfo  {journal}
  {J. Chem. Phys.}\ }\textbf {\bibinfo {volume} {68}},\ \bibinfo {pages}
  {2134--2139} (\bibinfo {year} {1978})}\BibitemShut {NoStop}%
\bibitem [{\citenamefont {Humphrey}, \citenamefont {Dalke},\ and\ \citenamefont
  {Schulten}(1996)}]{vmd}%
  \BibitemOpen
  \bibfield  {author} {\bibinfo {author} {\bibfnamefont {W.}~\bibnamefont
  {Humphrey}}, \bibinfo {author} {\bibfnamefont {A.}~\bibnamefont {Dalke}},\
  and\ \bibinfo {author} {\bibfnamefont {K.}~\bibnamefont {Schulten}},\
  }\bibfield  {title} {\enquote {\bibinfo {title} {Vmd: Visual molecular
  dynamics},}\ }\href@noop {} {\bibfield  {journal} {\bibinfo  {journal} {J.
  Molec. Graphics}\ }\textbf {\bibinfo {volume} {14}},\ \bibinfo {pages}
  {33--38} (\bibinfo {year} {1996})}\BibitemShut {NoStop}%
\bibitem [{\citenamefont {Batchelor}(1972)}]{batchelor:jfm:1972}%
  \BibitemOpen
  \bibfield  {author} {\bibinfo {author} {\bibfnamefont {G.~K.}\ \bibnamefont
  {Batchelor}},\ }\bibfield  {title} {\enquote {\bibinfo {title} {Sedimentation
  in a dilute dispersion of spheres},}\ }\href
  {https://doi.org/10.1017/S0022112072001399} {\bibfield  {journal} {\bibinfo
  {journal} {J. Fluid Mech.}\ }\textbf {\bibinfo {volume} {52}},\ \bibinfo
  {pages} {245–268} (\bibinfo {year} {1972})}\BibitemShut {NoStop}%
\bibitem [{\citenamefont {Ladd}(1990)}]{ladd:1990}%
  \BibitemOpen
  \bibfield  {author} {\bibinfo {author} {\bibfnamefont {A.~J.~C.}\
  \bibnamefont {Ladd}},\ }\bibfield  {title} {\enquote {\bibinfo {title}
  {Hydrodynamic transport coefficients of random dispersions of hard
  spheres},}\ }\href {https://doi.org/10.1063/1.458830} {\bibfield  {journal}
  {\bibinfo  {journal} {J. Chem. Phys.}\ }\textbf {\bibinfo {volume} {93}},\
  \bibinfo {pages} {3484--3494} (\bibinfo {year} {1990})}\BibitemShut {NoStop}%
\bibitem [{\citenamefont {Brady}\ and\ \citenamefont
  {Durlofsky}(1988)}]{brady:pf:1988}%
  \BibitemOpen
  \bibfield  {author} {\bibinfo {author} {\bibfnamefont {J.~F.}\ \bibnamefont
  {Brady}}\ and\ \bibinfo {author} {\bibfnamefont {L.~J.}\ \bibnamefont
  {Durlofsky}},\ }\bibfield  {title} {\enquote {\bibinfo {title} {The
  sedimentation rate of disordered suspensions},}\ }\href
  {https://doi.org/10.1063/1.866808} {\bibfield  {journal} {\bibinfo  {journal}
  {Phys. Fluids}\ }\textbf {\bibinfo {volume} {31}},\ \bibinfo {pages}
  {717--727} (\bibinfo {year} {1988})}\BibitemShut {NoStop}%
\bibitem [{\citenamefont {Banchio}\ and\ \citenamefont
  {Nägele}(2008)}]{banchio:2008}%
  \BibitemOpen
  \bibfield  {author} {\bibinfo {author} {\bibfnamefont {A.~J.}\ \bibnamefont
  {Banchio}}\ and\ \bibinfo {author} {\bibfnamefont {G.}~\bibnamefont
  {Nägele}},\ }\bibfield  {title} {\enquote {\bibinfo {title} {Short-time
  transport properties in dense suspensions: From neutral to charge-stabilized
  colloidal spheres},}\ }\href {https://doi.org/10.1063/1.2868773} {\bibfield
  {journal} {\bibinfo  {journal} {J. Chem. Phys.}\ }\textbf {\bibinfo {volume}
  {128}},\ \bibinfo {pages} {104903} (\bibinfo {year} {2008})}\BibitemShut
  {NoStop}%
\bibitem [{\citenamefont {Whitmer}\ and\ \citenamefont
  {Luijten}(2010)}]{whitmer:2010}%
  \BibitemOpen
  \bibfield  {author} {\bibinfo {author} {\bibfnamefont {J.~K.}\ \bibnamefont
  {Whitmer}}\ and\ \bibinfo {author} {\bibfnamefont {E.}~\bibnamefont
  {Luijten}},\ }\bibfield  {title} {\enquote {\bibinfo {title} {Fluid–solid
  boundary conditions for multiparticle collision dynamics},}\ }\href
  {https://doi.org/10.1088/0953-8984/22/10/104106} {\bibfield  {journal}
  {\bibinfo  {journal} {J. Phys.: Condens. Matter}\ }\textbf {\bibinfo {volume}
  {22}},\ \bibinfo {pages} {104106} (\bibinfo {year} {2010})}\BibitemShut
  {NoStop}%
\bibitem [{\citenamefont {Zantop}\ and\ \citenamefont
  {Stark}(2021)}]{zantop:2021}%
  \BibitemOpen
  \bibfield  {author} {\bibinfo {author} {\bibfnamefont {A.~W.}\ \bibnamefont
  {Zantop}}\ and\ \bibinfo {author} {\bibfnamefont {H.}~\bibnamefont {Stark}},\
  }\bibfield  {title} {\enquote {\bibinfo {title} {Multi-particle collision
  dynamics with a non-ideal equation of state. i},}\ }\href
  {https://doi.org/10.1063/5.0037934} {\bibfield  {journal} {\bibinfo
  {journal} {J. Chem. Phys.}\ }\textbf {\bibinfo {volume} {154}},\ \bibinfo
  {pages} {024105} (\bibinfo {year} {2021})}\BibitemShut {NoStop}%
\bibitem [{\citenamefont {{Lamura, A.}}\ \emph {et~al.}(2001)\citenamefont
  {{Lamura, A.}}, \citenamefont {{Gompper, G.}}, \citenamefont {{Ihle, T.}},\
  and\ \citenamefont {{Kroll, D. M.}}}]{lamura:2001}%
  \BibitemOpen
  \bibfield  {author} {\bibinfo {author} {\bibnamefont {{Lamura, A.}}},
  \bibinfo {author} {\bibnamefont {{Gompper, G.}}}, \bibinfo {author}
  {\bibnamefont {{Ihle, T.}}},\ and\ \bibinfo {author} {\bibnamefont {{Kroll,
  D. M.}}},\ }\bibfield  {title} {\enquote {\bibinfo {title} {Multi-particle
  collision dynamics: Flow around a circular and a square cylinder},}\ }\href
  {https://doi.org/10.1209/epl/i2001-00522-9} {\bibfield  {journal} {\bibinfo
  {journal} {Europhys. Lett.}\ }\textbf {\bibinfo {volume} {56}},\ \bibinfo
  {pages} {319--325} (\bibinfo {year} {2001})}\BibitemShut {NoStop}%
\bibitem [{\citenamefont {Bolintineanu}\ \emph {et~al.}(2012)\citenamefont
  {Bolintineanu}, \citenamefont {Lechman}, \citenamefont {Plimpton},\ and\
  \citenamefont {Grest}}]{bolintineanu:2012}%
  \BibitemOpen
  \bibfield  {author} {\bibinfo {author} {\bibfnamefont {D.~S.}\ \bibnamefont
  {Bolintineanu}}, \bibinfo {author} {\bibfnamefont {J.~B.}\ \bibnamefont
  {Lechman}}, \bibinfo {author} {\bibfnamefont {S.~J.}\ \bibnamefont
  {Plimpton}},\ and\ \bibinfo {author} {\bibfnamefont {G.~S.}\ \bibnamefont
  {Grest}},\ }\bibfield  {title} {\enquote {\bibinfo {title} {No-slip boundary
  conditions and forced flow in multiparticle collision dynamics},}\ }\href
  {https://doi.org/10.1103/PhysRevE.86.066703} {\bibfield  {journal} {\bibinfo
  {journal} {Phys. Rev. E}\ }\textbf {\bibinfo {volume} {86}},\ \bibinfo
  {pages} {066703} (\bibinfo {year} {2012})}\BibitemShut {NoStop}%
\bibitem [{\citenamefont {Trokhymchuk}\ \emph {et~al.}(2005)\citenamefont
  {Trokhymchuk}, \citenamefont {Nezbeda}, \citenamefont {Jirsák},\ and\
  \citenamefont {Henderson}}]{trokhymchuk:2005}%
  \BibitemOpen
  \bibfield  {author} {\bibinfo {author} {\bibfnamefont {A.}~\bibnamefont
  {Trokhymchuk}}, \bibinfo {author} {\bibfnamefont {I.}~\bibnamefont
  {Nezbeda}}, \bibinfo {author} {\bibfnamefont {J.}~\bibnamefont {Jirsák}},\
  and\ \bibinfo {author} {\bibfnamefont {D.}~\bibnamefont {Henderson}},\
  }\bibfield  {title} {\enquote {\bibinfo {title} {{Hard-sphere radial
  distribution function again}},}\ }\href {https://doi.org/10.1063/1.1979488}
  {\bibfield  {journal} {\bibinfo  {journal} {J. Chem. Phys.}\ }\textbf
  {\bibinfo {volume} {123}},\ \bibinfo {pages} {024501} (\bibinfo {year}
  {2005})}\BibitemShut {NoStop}%
\bibitem [{\citenamefont {Anderson}, \citenamefont {Eric~Irrgang},\ and\
  \citenamefont {C.~Glotzer}(2016)}]{anderson:cpc:2016}%
  \BibitemOpen
  \bibfield  {author} {\bibinfo {author} {\bibfnamefont {J.~A.}\ \bibnamefont
  {Anderson}}, \bibinfo {author} {\bibfnamefont {M.}~\bibnamefont
  {Eric~Irrgang}},\ and\ \bibinfo {author} {\bibfnamefont {S.}~\bibnamefont
  {C.~Glotzer}},\ }\bibfield  {title} {\enquote {\bibinfo {title} {Scalable
  metropolis monte carlo for simulation of hard shapes},}\ }\href
  {https://doi.org/https://doi.org/10.1016/j.cpc.2016.02.024} {\bibfield
  {journal} {\bibinfo  {journal} {Comput. Phys. Commun.}\ }\textbf {\bibinfo
  {volume} {204}},\ \bibinfo {pages} {21--30} (\bibinfo {year}
  {2016})}\BibitemShut {NoStop}%
\bibitem [{\citenamefont {te~Vrugt}\ and\ \citenamefont
  {Wittkowski}(2022)}]{tevrugt:2023}%
  \BibitemOpen
  \bibfield  {author} {\bibinfo {author} {\bibfnamefont {M.}~\bibnamefont
  {te~Vrugt}}\ and\ \bibinfo {author} {\bibfnamefont {R.}~\bibnamefont
  {Wittkowski}},\ }\bibfield  {title} {\enquote {\bibinfo {title} {Perspective:
  New directions in dynamical density functional theory},}\ }\href
  {https://doi.org/10.1088/1361-648X/ac8633} {\bibfield  {journal} {\bibinfo
  {journal} {J. Phys.: Condens.Matter}\ }\textbf {\bibinfo {volume} {35}},\
  \bibinfo {pages} {041501} (\bibinfo {year} {2022})}\BibitemShut {NoStop}%
\bibitem [{\citenamefont {de~las Heras}\ \emph {et~al.}(2023)\citenamefont
  {de~las Heras}, \citenamefont {Zimmermann}, \citenamefont {Sammüller},
  \citenamefont {Hermann},\ and\ \citenamefont {Schmidt}}]{delasheras:2023}%
  \BibitemOpen
  \bibfield  {author} {\bibinfo {author} {\bibfnamefont {D.}~\bibnamefont
  {de~las Heras}}, \bibinfo {author} {\bibfnamefont {T.}~\bibnamefont
  {Zimmermann}}, \bibinfo {author} {\bibfnamefont {F.}~\bibnamefont
  {Sammüller}}, \bibinfo {author} {\bibfnamefont {S.}~\bibnamefont
  {Hermann}},\ and\ \bibinfo {author} {\bibfnamefont {M.}~\bibnamefont
  {Schmidt}},\ }\bibfield  {title} {\enquote {\bibinfo {title} {Perspective:
  How to overcome dynamical density functional theory},}\ }\href
  {https://doi.org/10.1088/1361-648X/accb33} {\bibfield  {journal} {\bibinfo
  {journal} {J. Phys.: Condens. Matter}\ }\textbf {\bibinfo {volume} {35}},\
  \bibinfo {pages} {271501} (\bibinfo {year} {2023})}\BibitemShut {NoStop}%
\bibitem [{\citenamefont {Schmidt}\ and\ \citenamefont
  {Brader}(2013)}]{schmidt:2013}%
  \BibitemOpen
  \bibfield  {author} {\bibinfo {author} {\bibfnamefont {M.}~\bibnamefont
  {Schmidt}}\ and\ \bibinfo {author} {\bibfnamefont {J.~M.}\ \bibnamefont
  {Brader}},\ }\bibfield  {title} {\enquote {\bibinfo {title} {{Power
  functional theory for Brownian dynamics}},}\ }\href
  {https://doi.org/10.1063/1.4807586} {\bibfield  {journal} {\bibinfo
  {journal} {J. Chem. Phys.}\ }\textbf {\bibinfo {volume} {138}},\ \bibinfo
  {pages} {214101} (\bibinfo {year} {2013})}\BibitemShut {NoStop}%
\bibitem [{\citenamefont {Schmidt}(2022)}]{schmidt:rmp:2022}%
  \BibitemOpen
  \bibfield  {author} {\bibinfo {author} {\bibfnamefont {M.}~\bibnamefont
  {Schmidt}},\ }\bibfield  {title} {\enquote {\bibinfo {title} {Power
  functional theory for many-body dynamics},}\ }\href@noop {} {\bibfield
  {journal} {\bibinfo  {journal} {Rev. Mod. Phys.}\ }\textbf {\bibinfo {volume}
  {94}},\ \bibinfo {pages} {015007} (\bibinfo {year} {2022})}\BibitemShut
  {NoStop}%
\end{thebibliography}%

\end{document}


\title{Supplementary material for ``Exploring the role of hydrodynamic interactions in spherically-confined drying colloidal suspensions''}

\author{Mayukh Kundu}
\thanks{These authors contributed equally.}
\affiliation{Department of Chemical Engineering, Auburn University, Auburn, AL 36849, USA}

\author{Kritika Kritika}
\thanks{These authors contributed equally.}
\affiliation{Leibniz-Institut f{\"u}r Polymerforschung Dresden e.V., Hohe Stra{\ss}e 6, 01069 Dresden, Germany}
\affiliation{Institut f{\"u}r Theoretische Physik, Technische Universit{\"a}t Dresden, 01069 Dresden, Germany}

\author{Yashraj M. Wani}
\affiliation{Institute of Physics, Johannes Gutenberg University Mainz, Staudingerweg 7, 55128 Mainz, Germany}

\author{Arash Nikoubashman}
\email{anikouba@ipfdd.de}
\affiliation{Leibniz-Institut f{\"u}r Polymerforschung Dresden e.V., Hohe Stra{\ss}e 6, 01069 Dresden, Germany}
\affiliation{Institut f{\"u}r Theoretische Physik, Technische Universit{\"a}t Dresden, 01069 Dresden, Germany}

\author{Michael P. Howard}
\email{mphoward@auburn.edu}
\affiliation{Department of Chemical Engineering, Auburn University, Auburn, AL 36849, USA}

\maketitle

\clearpage
\section{Particle distributions at additional initial volume fractions}
\begin{figure}[!h]
    \centering
    \includegraphics{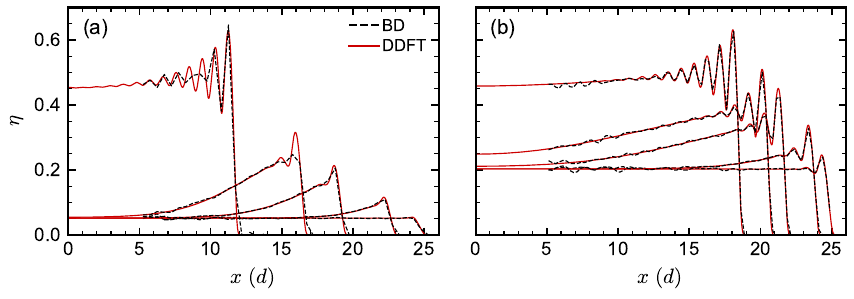}
    \caption{Same as Fig.~2(b) for (a) $\eta_0 = 0.05$ and (b) $\eta_0 = 0.20$.}
\end{figure}

\begin{figure}[!h]
    \centering
    \includegraphics{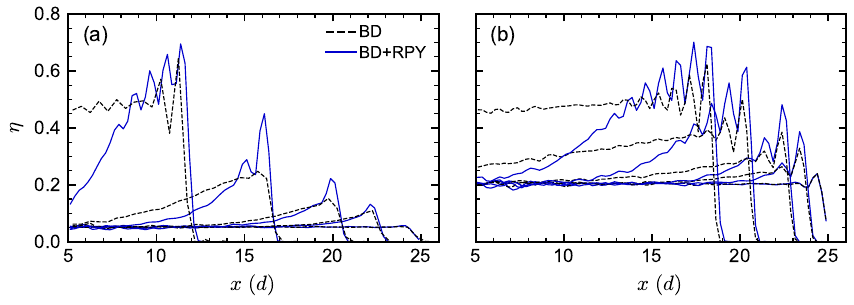}
    \caption{Same as Fig.~3(a) for (a) $\eta_0 = 0.05$ and (b) $\eta_0 = 0.20$.}
\end{figure}

\clearpage
\section{Multiparticle collision dynamics}
\begin{figure}[!h]
    \centering
    \includegraphics{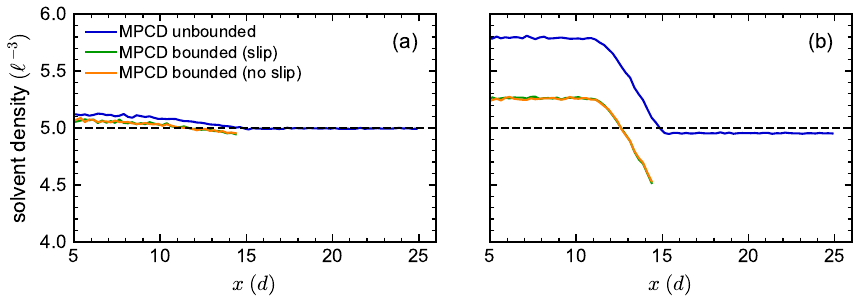}
    \caption{MPCD solvent density as a function of radial distance $x$ from center of droplet at the end of drying with (a) ${\rm Pe} = 10$ and (b) ${\rm Pe} = 100$ for $\eta_0 = 0.10$ and unbounded, slip, and no-slip boundary conditions on the solvent at the droplet's liquid--vapor interface. The dashed line indicates the desired uniform solvent density.}
\end{figure}

\begin{figure}[!h]
    \centering
    \includegraphics{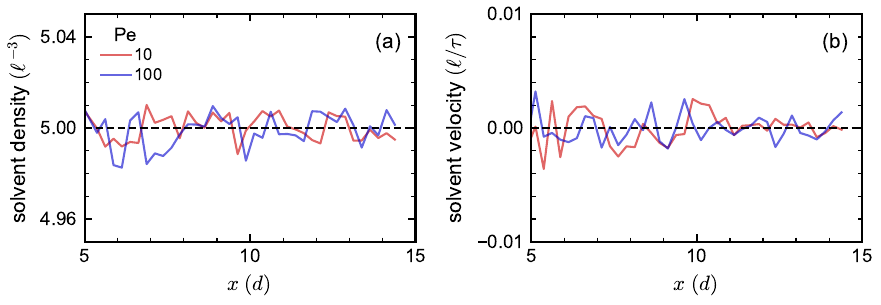}
    \caption{MPCD solvent (a) density and (b) radial velocity as a function of radial distance $x$ from center of droplet at the end of drying with ${\rm Pe} = 10$ and $100$ for a pure solvent with slip boundary conditions at the droplet's liquid--vapor interface. The dashed lines indicate the expected uniform solvent density and zero solvent velocity for an incompressible fluid.}
\end{figure}

\begin{figure}[!h]
    \centering
    \includegraphics{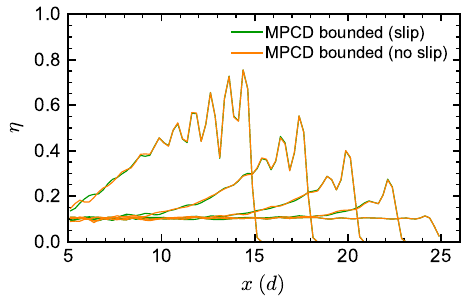}
    \caption{Same as Fig.~4 but now comparing MPCD simulation with a slip boundary condition and a no-slip boundary condition on the solvent at the droplet's liquid--vapor interface.}
\end{figure}

\clearpage
\section{Dynamic density functional theory}
\subsection{Integration bounds for $M^{(2)}$ and $j^{(2)}$}
\begin{figure}[!h]
    \centering
    \includegraphics{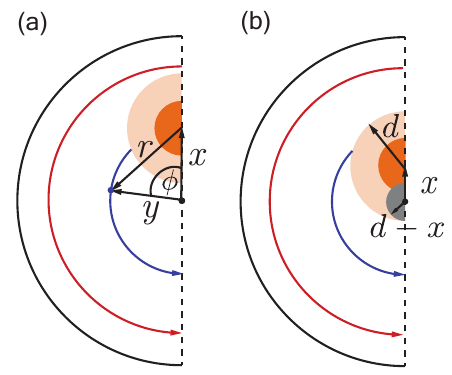}
    \caption{Schematic of integration paths for $M^{(2)}$. The red line represents the case where $|x-y| > d$ and there are no restrictions on $\phi$, while the blue line represents the case where $|x-y| \leq d$ and there are restrictions on $\phi$. The solid orange region represents the hard sphere at $x$, and the shaded orange region represents the excluded region around it. In (b), the gray region shows where $y$ has a lower bound of $d-x$ because there is overlap for all $\phi$ (i.e., the gray region is fully enclosed by the shaded orange region).}
    \label{fig:integral_bounds}
\end{figure}

The integration bounds on $\cos \phi$ in Eq.~(25) should be adjusted based on $x$ and $y$ because $g$ is zero when two hard spheres overlap. Figure \ref{fig:integral_bounds} illustrates both an integration path where there are no restrictions on $\phi$ because $|x-y| > d$ (in red) and an integration path where there are restrictions on $\phi$ because $|x-y| \le d$ (in blue). In the latter case, the lower bound on $\phi$ (upper bound on $\cos \phi$) is computed using the law of cosines. Additionally, Fig.~\ref{fig:integral_bounds}(b) illustrates how the lower integration bound on $y$ in Eq.~(24) must be greater than $d-x$ if $x < d$ because there is overlap for all values of $\phi$ (gray region).

\subsection{DDFT with pairwise HIs for density-dependent $g$}
\begin{figure}[!h]
    \centering
    \includegraphics{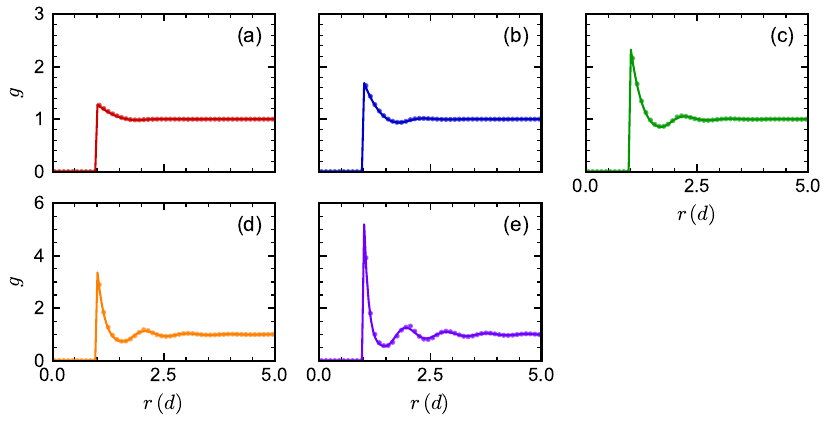}
    \caption{Bulk radial distribution function $g$ for volume fractions (a) 0.10, (b) 0.20, (c) 0.30, (d) 0.40, and (e) 0.50 calculated using hard-sphere Monte Carlo simulation (points) and the form given in Ref.~93 (lines).}
    \label{fig:bulk_rdf}
\end{figure}

We first confirmed that the density-dependent form of $g$ proposed in Ref.~93 accurately predicts the radial distribution function for a bulk hard-sphere suspension. We conducted bulk equilibrium hard particle Monte Carlo simulations at five different volume fractions (0.10, 0.20, 0.30, 0.40, and 0.50), running two independent simulations for each volume fraction. Particles were first placed randomly in a cubic box of side length $70\,d$ without overlap. Then HOOMD-blue (version 4.7.0) was used to perform the simulations with an maximum particle displacement of $0.1\,d$. A simulation step was defined as accepting two trial moves per particle, each simulation was run for $5 \times 10^7$ steps, and configurations were saved every $5 \times 10^5$ steps for analysis. The simulations and predictions were found to be in excellent agreement (Fig.~\ref{fig:bulk_rdf}). We then moved on to incorporate the density-dependent form of $g$ from Ref.~93 in Eq.~(25). This modification led to two primary differences from the approach we used for the dilute $g$: (1) $M^{(2)}$ was no longer guaranteed to have a highly localized ($|x-y| < d$) support and (2) $M^{(2)}$ no longer had a simple analytical form.

\begin{figure}[!h]
    \centering
    \includegraphics{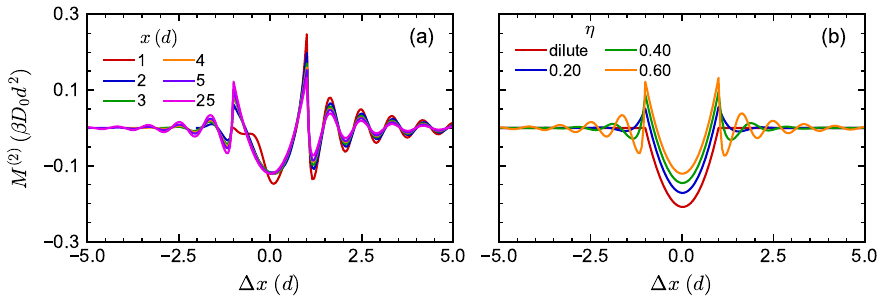}
    \caption{Variation in $M^{(2)}$ with respect to $\Delta x$ for (a) different values of $x$ at mean density $\bar\rho$ consistent with volume fraction 0.60 and (b) different volume fractions at $x = 5\,d$. In (a), only points for which $y \ge 0\,d$ are shown.}
    \label{fig:m2_nondilute}
\end{figure}

Difference 1 nominally requires integration over all $y$ to evaluate the flux $j^{(2)}$ using Eq.~(24), which is undesirable because of computational cost, so we first investigated whether we could reasonably truncate this integration. We anticipated $M^{(2)}$ might be most difficult to truncate for denser suspensions, for which longer ranged oscillations in particle correlations develop. Accordingly, we reexpressed $y$ using $\Delta x = y - x$ and inspected the behavior of $M^{(2)}$ as a function of $\Delta x$ at a volume fraction of $0.60$ and different values of $x$ [Fig.~\ref{fig:m2_nondilute}(a)]. This volume fraction was selected as an upper bound because it is close to that of randomly close packed spheres; although it is somewhat outside the range of volume fractions recommended in Ref.~93, we confirmed that $g$ remained nonnegative and did not show any obvious artifacts. For most values of $x$, $M^{(2)}$ exhibited a sharp peak near $\Delta x = \pm 1\,d$ (the point where the dilute $M^{(2)}$ becomes zero) then asymmetric decaying oscillations. It also initially had a dependence on $x$ but became essentially independent of $x$ for $x \ge 5\,d$. We additionally confirmed that our expectation that more significant oscillations occurred for larger volume fraction [Fig.~\ref{fig:m2_nondilute}(b)]. Overall, the observed decay in the oscillations suggested that truncation might be reasonable.

To systematically select a value of $\Delta x$ at which to truncate integration of $M^{(2)}$, we numerically computed
\begin{equation}
I = \int_{x-\Delta x}^{x+\Delta x} \d{y} M^{(2)}(x, y; \bar \rho).
\end{equation}
This integral is a reasonable proxy for $j^{(2)}$ for sufficiently small variation in $\rho(x)$. We compared $I$ to the reference integral $I_\infty$ obtained by numerically integrating $M^{(2)}$ for $0 \le y < \infty$. We then analyzed the difference $\Delta I = I - I_\infty$ as a function of $\Delta x$ for different values of $x$ with $\bar\rho$ again consistent with volume fraction 0.60. As expected, the relative error $|\Delta I/I_\infty|$ tended to decrease toward zero as $\Delta x$ increased, with most of the decrease occurring before $\Delta x = 2\, d$. Importantly, we found that when $\Delta x = 3\,d$, the relative error was less than 5\% and only about 2\% for most values of $x$. We accordingly chose to evaluate the integral in Eq.~(25) for $y$ between $x \pm 3\,d$ in order to balance accuracy with computational cost.
\begin{figure}
    \centering
    \includegraphics{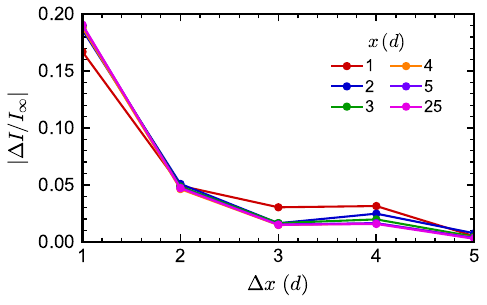}
    \caption{Relative error $\Delta I/I_\infty$ as a function of truncation length $\Delta x$ for different values of $x$ at average density $\bar\rho$ consistent with volume fraction 0.60.}
\end{figure}

Difference 2 nominally requires us to numerically recalculate $M^{(2)}(x,y;\bar\rho)$ for each mesh point as the density evolves in our DDFT simulations, but the cost of repeatedly doing so would be prohibitive. To circumvent this issue, we instead precomputed $M^{(2)}$ at selected values of $x$, $\Delta x$, and $\bar\rho$ and employed a linear interpolation scheme, as described in the main text. As a minor detail that is only relevant for values of $x$ close to the center of the droplet, we use a value of zero in our interpolation for any combinations of $x$ and $\Delta x$ that lie outside the integration bounds on $y$.

\subsection{Force profile}
\begin{figure}[!h]
    \centering
    \includegraphics{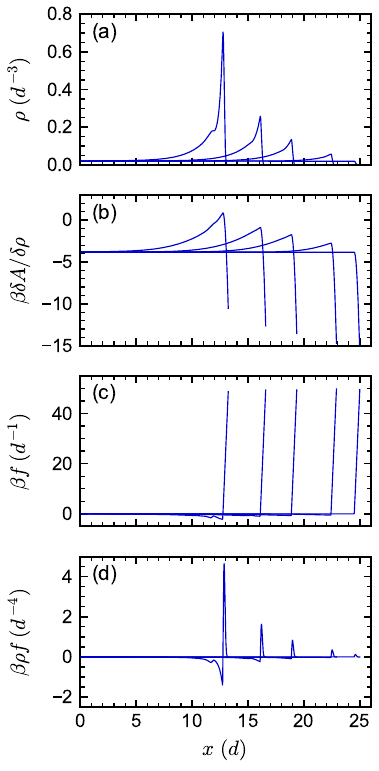}
    \caption{(a) Density $\rho$ (b) functional derivative of free energy $\delta A/\delta \rho$ (c) force $f$, and (d) force density $\rho f$ calculated using DDFT for the volume fraction profiles computed using DDFT with RPY HIs and dilute pair correlations in Fig.~6.}
\end{figure}